\documentclass[11pt,letterpaper]{article} 

\usepackage[margin=1in]{geometry} 
\usepackage{setspace}
\usepackage{fullpage}
\usepackage{graphicx}

\usepackage{microtype}

\usepackage{amssymb}
\usepackage{amsmath}
\usepackage{amsthm}

\usepackage{xspace}
\usepackage{color}

\usepackage{todonotes}
\usepackage{hyperref}

\newtheorem{theorem}{Theorem}
\newtheorem{definition}[theorem]{Definition}
\newtheorem{lemma}[theorem]{Lemma}
\newtheorem{corollary}[theorem]{Corollary}

\newenvironment{denseitems}{\list{$\bullet$}{\itemsep6pt\parsep-5pt}}{\endlist}

\let\leq\leqslant
\let\geq\geqslant

\graphicspath{{figures/}}

\title{The Containment Polytope of a Triangle and a Convex Polygon}

\author{Yoonsung Choi (KAIST), Sunghee Choi (KAIST)}

\newcommand{\Ptheta}{\mathcal{P}(\theta)}
\newcommand{\Ttheta}{\mathcal{T}(\theta)}

\newcommand{\atheta}{a_{\theta}}
\newcommand{\btheta}{b_{\theta}}
\newcommand{\ctheta}{c_{\theta}}

\newcommand{\PP}{\mathcal{P}}

\newcommand{\TT}{\mathcal{T}}

\newcommand{\Reals}{\mathbb{R}}
\newcommand{\CC}{\mathcal{C}}
\newcommand{\FF}{\mathcal{F}}
\newcommand{\silhouette}{\Lambda}

\newcommand{\EE}{\mathcal{E}}
\newcommand{\Etheta}{\mathcal{E}(\theta)}

\DeclareMathOperator{\polylog}{polylog}

\usepackage{hyperref}
\hypersetup{%
  breaklinks,%
  colorlinks=true,%
  linkcolor=[rgb]{0.45,0.0,0.0},%
  citecolor=[rgb]{0,0,0.45}
}

\makeatletter
\partopsep\z@ \textfloatsep 10pt plus 1pt minus 4pt
\def\section{\@startsection {section}{1}{\z@}%
  {-3.5ex plus -1ex
    minus -.2ex}{2.3ex plus .2ex}{\large\bf}}
\def\subsection{\@startsection{subsection}{2}%
  {\z@}{-3.25ex plus
    -1ex minus -.2ex}{1.5ex plus .2ex}{\normalsize\bf}}
\def\@fnsymbol#1{\ensuremath{\ifcase#1\or 1\or 2\or 3\or 4\or
    5\or 6\or 7 \or 8\ or 9 \or 10\or 11 \else\@ctrerr\fi}}
\makeatother

\begin{document}

\maketitle

\begin{abstract}
  The \emph{containment polytope}~$\CC$ for a triangle~$P$ and a
  convex $n$-gon~$Q$ is a four-dimensional polytope representing the
  similarity transforms (translation, rotation, scaling) that
  place~$P$ inside~$Q$.  While it is known that the complexity
  of~$\CC$ and its silhouette $\Lambda$,
  which represents all the largest placements of $P$
  inside $Q$ for all orientations, can be quadratic in~$n$, we show that
  the silhouette can be partitioned into a \emph{linear} number
  of \emph{planar} convex chains and steps in~$O(n \log n)$ time.
  
  Based on the linear number of convex chains and steps,
  we can compute the complexity of silhouette~$\Lambda$
  in $O(n \log n)$ time and query the largest placement of~$P$ inside $Q$
  for a given orientation in $O(\log n)$ time.
  We explain geometrically how the triangle~$P$
  rotates while continuously scaling and translating to be maximal
  inside~$Q$, leading to a parameterized algorithm for
  containment problems.
\end{abstract} 

\section{Introduction}
\label{sec:intro}

For a convex $m$-gon~$P$ and a convex $n$-gon~$Q$ in the plane, it is
a natural question to ask (1)~whether $P$ can be placed inside~$Q$
allowing translation and rotation, or, more generally, (2)~what is the
largest similar copy of~$P$ that can be placed inside~$Q$ allowing
translation and rotation.

In 1983, Chazelle~\cite{chazelle1983polygon} gave an $O(mn^{2})$-time
algorithm for the first question.  When $P$ is an equilateral triangle
or a square, DePano, Ke and O'Rourke~\cite{depano1987finding} gave an
$O(n^{2})$-time algorithm for the second question.  When~$P$ is a
general triangle, Sharir and Toledo~\cite{sharir1994external} used
parametric search to obtain an algorithm for the second question with running
time~$O(n^{2} \log^{2} n)$.

In 1998, Agarwal, Amenta and Sharir~\cite{agarwal1998largest} gave
an~$O(mn^{2} \log n)$-time algorithm to compute the set of \emph{all
possible} placements of~$P$ inside~$Q$ under translation, rotation,
and scaling.  They followed Baird~\cite{baird}, to represent all
similarity transforms (that is, combinations of scaling, rotation, and
translation) by quadruples~$(s, t, u, v)$ such that~$(s, t) = (\lambda
\cos \theta, \lambda \sin \theta)$ represent scaling~$\lambda$ and
rotation~$\theta$, while~$(u, v)$ represents the translation.
For~$(s, t, u, v) \in \Reals^{4}$, the corresponding similarity
transform is the affine map~$(x, y) \mapsto (sx - ty + u, tx + sy +
v)$.  They define the set~$\CC \subset \Reals^{4}$ as the set of all
similarity transforms~$\pi$ such that~$\pi P \subseteq Q$.

We have~$\pi P \subseteq Q$ if and only if all~$m$ vertices of~$P$ lie
inside all the~$n$ halfplanes that define the convex polygon~$Q$.
For vertex $v_i = (x_{i}, y_{i})$ of~$P$ and halfplane
$h_j: a_{j} x + b_{j} y \leq 1$ of~$Q$, we have $\pi v_i \in h_j$ if and
only if
\begin{align*}
 a_{j}(sx_{i} - ty_{i} + u) + b_{j}(tx_{i} + sy_{i} + v) \leq 1,
\end{align*}
which is a linear inequality in the four variables~$(s, t, u, v)$.  It
follows that the set~$\CC$ is the four-dimensional convex polytope
obtained as the intersection of~$mn$ half-spaces.  We call~$\CC$ the
\emph{containment polytope} of~$P$ and~$Q$. The authors
of~\cite{agarwal1998largest} show that the complexity of the
containment polytope is~$\Theta(mn^{2})$.

If we project~$\CC$ onto the $st$-plane, we obtain a two-dimensional
convex polygon~$\FF$. Scaling and rotation are the polar coordinates
of this plane, which means that~$P$ can be scaled by~$\lambda$ and
rotated by~$\theta$ and placed in~$Q$ if and only if~$(s, t) =
(\lambda \cos \theta, \lambda \sin \theta) \in \FF$.

For an angle~$\theta$, let us define~$P_{\theta}$ as~$P$ rotated
counter-clockwise by angle~$\theta$ around the origin.  We next define
$\lambda(\theta)$ as the largest scaling factor such that
$\lambda(\theta)P_\theta$ can be placed inside~$Q$.  From the above,
it follows that the boundary of~$\FF$ is a representation of the
function~$\theta \mapsto \lambda(\theta)$.  The maximum is necessarily
attained at a vertex of~$\FF$.  The authors
of~\cite{agarwal1998largest} show how to compute the vertices
of~$\CC$, and therefore of~$\FF$, in~$O(mn^{2} \log n)$ time, thus
solving problem~(2).  They also construct convex polygons~$P$ and~$Q$
such that the containment polytope has~$\Theta(m n^{2})$ complexity.

This lower bound was strengthened by the authors
of~\cite{bae2018reverse}, who, for every~$m$ and~$n$ construct convex
polygons~$P$ and~$Q$ such that the \emph{projection}~$\FF$ has
complexity~$\Theta(m n^{2})$.  For~$m=3$, $P$~can even be chosen as
the equilateral triangle.  This implies that the function~$\theta
\mapsto \lambda(\theta)$ does have complexity~$\Theta(m n^{2})$.
Various
authors~\cite{chazelle1983polygon,lee2021largest,agarwal1998largest,bae2018reverse}
have taken this as an indication that solving problem~(2) will require
time~$\Omega(m n^{2})$, even in the case of triangular~$P$.

Nevertheless, in 2024, Chan and Hair~\cite{chan2024convex} achieved a
break-through: they presented an $O(n \polylog n)$-time algorithm for
problem~(2) when $P$ is a triangle, making use of parametric search
and divide-and-conquer to avoid having to look at all vertices
of~$\FF$. They extend this to an
$O(m^{O(1/\varepsilon)}n^{1+\varepsilon})$-time algorithm (for any
$\varepsilon > 0$) when $P$ is a convex~$m$-gon

In some situations one is actually interested in other aspects of the
function~$\theta \mapsto \lambda(\theta)$.  The interest
of~\cite{bae2018reverse} was to answer a question paused by
P\'al~\cite{pal1921} in his work on Kakeya's problem: given two convex
shapes (not necessarily polygons)~$P$ and~$Q$ and knowing that for
every~$\theta$, $P_\theta$ can be placed in~$Q$, is it then true
that~$P$ can be rotated through~$360^{\circ}$ inside~$Q$?  The paper
shows that this is indeed true, and proves the following theorem
(which for the polygonal case easily follows from the description
using the containment polytope~$\CC$):
\begin{theorem}[\cite{bae2018reverse}]
  \label{thm:continuous}
  For any convex shapes $P$ and $Q$, there exists a continuous function
  $\tau : [0, 2\pi] \rightarrow \mathbb{R}^{2}$
  such that $\lambda(\theta)P_{\theta} + \tau(\theta) \subseteq Q$.
\end{theorem}

\begin{figure}
  \centerline{\includegraphics[width=0.5\textwidth]{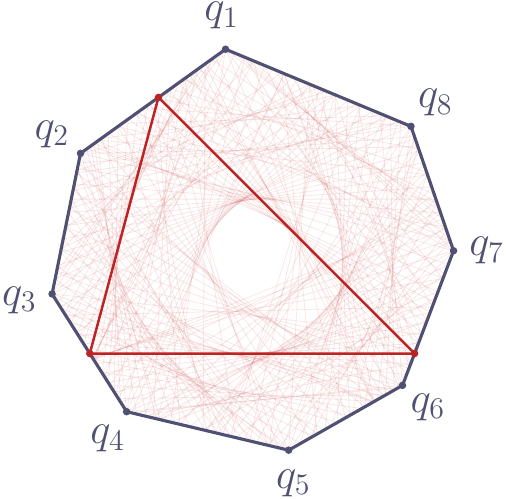}}
  \caption{The triangles $\lambda(\theta)P_{\theta} + \tau(\theta)$ as
    we rotate through~$360^{\circ}$.}
  \label{fig:silhouette}
\end{figure}

We can now formulate a third problem on the convex polygons~$P$
and~$Q$: (3) find the largest scaling~$\lambda$ of~$P$ such
that~$\lambda P$ can be rotated through~$360^{\circ}$ inside~$Q$.
Solving this problem requires finding the \emph{minimum}
of~$\lambda(\theta)$.

The function~$\theta \mapsto (\lambda(\theta), \tau(\theta))$
corresponds to the \emph{silhouette}~$\silhouette$ of the containment
polytope~$\CC$, that is, those edges of~$\CC$ that project onto the
edges of~$\FF$.  The silhouette forms a closed loop in~$\Reals^{4}$.

In this paper, we will study the geometry of the
triangle~$\lambda(\theta)P_{\theta} + \tau(\theta)$ as~$\theta$
rotates through~$360^{\circ}$, in other words, as we trace the
silhouette~$\silhouette$ (see Figure~\ref{fig:silhouette}).

Throughout the paper, we will assume that the containment
polytope~$\CC$ is in \emph{general position}. Specifically we require
that it has no vertical faces, and that its defining hyperplanes are
in general position: no three hyperplanes intersect in a plane, no
four in a line, and no five in a point.  The first condition ensures
that the silhouette~$\silhouette$ is indeed a closed polygonal chain
consisting of edges (and does not include two-dimensional faces).
This condition holds when~$Q$ has no parallel edges.  The second
condition ensures that at every placement of the
triangle~$\lambda(\theta)P_{\theta} + \tau(\theta)$ on an edge
of~$\silhouette$, we have exactly three incidences between vertices
of~$P$ and edges of~$Q$, and at every vertex of~$\silhouette$, we have
exactly four such incidences.  Geometrically, this assumption means
that there is no scaled placement of~$P$ in~$Q$ with all vertices on
the boundary of~$Q$ and two vertices on vertices of~$Q$.  Without
these two assumptions, our proofs would need additional technical
complications that we currently do not know how to handle.

To avoid confusion between the vertices and edges of~$P$ and~$Q$ on
one side and of~$\CC$ on the other side, we will call the edges of the
silhouette~$\silhouette$ \emph{steps}, and the vertices
of~$\silhouette$ \emph{pivots}.  We will also call the vertices of~$P$
\emph{corners} to avoid confusion with the vertices of~$Q$.

For an angle~$\theta$, let~$\Ptheta$ be the unique largest placement
of~$P_{\theta}$ in~$Q$, that is~$\Ptheta = \lambda(\theta)P_{\theta} +
\tau(\theta)$, and let $\atheta, \btheta, \ctheta$ be the vertices $a,
b, c$, respectively, of~$\Ptheta$.

A \emph{perfect rotation} from $\theta_{1}$ to $\theta_{2}$ is the
set $\PP[\theta_{1}, \theta_{2}] = \{ \Ptheta\ |\ \theta_{1} \leq
\theta \leq \theta_{2}\}$.
A perfect rotation corresponds to
a subpath of the silhouette~$\silhouette$.
We naturally extend this notation to open and half-open intervals,
$\PP(\theta_{1}, \theta_{2})$, $\PP[\theta_{1}, \theta_{2})$,
$\PP(\theta_{1}, \theta_{2}]$, by excluding the corresponding endpoints.
We call $\PP[0, 2\pi]$ (the entire silhouette)
the \emph{global perfect rotation}.  Throughout
the paper, if we say $\Ptheta$ \emph{rotates}, this means that
$\Ptheta$ undergoes an \emph{perfect rotation} inside~$Q$ in
counter-clockwise direction.

For a given pivot $\sigma$, we define the \emph{next pivot} $\sigma'$
of $\sigma$ as the smallest pivot strictly larger than $\sigma$.
Then, the perfect rotation $\PP[\sigma, \sigma']$ corresponds to
a step of $\Lambda$.

\begin{figure}
  \centerline{\includegraphics[width=0.7\textwidth]{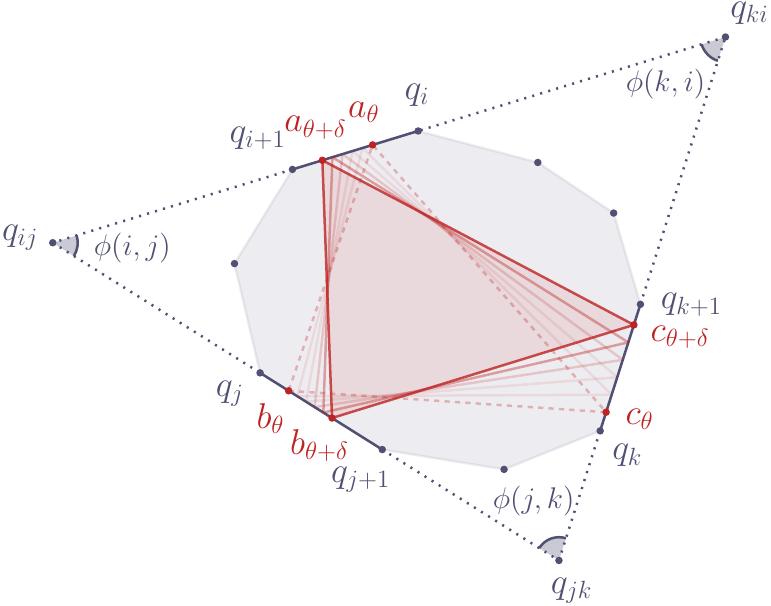}}
  \caption{The perfect rotation when the center lies inside $Q$.}
  \label{fig:inscribed_rotation_three_in}
\end{figure}

\begin{figure}
  \centerline{\includegraphics[width=1.0\textwidth]{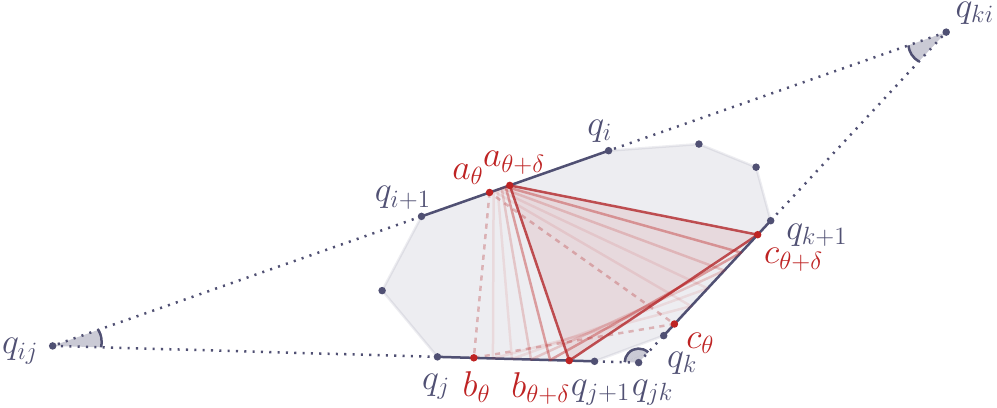}}
  \caption{The perfect rotation when the center lies outside $Q$.}
  \label{fig:inscribed_rotation_three_out}
\end{figure}

We start by showing that along a single step of~$\silhouette$, the
triangle~$P$ performs a perfect rotation with center $p$ (See
Figures~\ref{fig:inscribed_rotation_three_in}
and~\ref{fig:inscribed_rotation_three_out} for illustrations).
\begin{theorem}
  \label{thm:perfect}
  Let $P$ be a triangle and~$Q$ a convex polygon in general position,
  and consider a single step~$\PP[\sigma, \sigma']$
  of the silhouette~$\silhouette$.
  While the triangle $\Ptheta$ moves along this step,
  it performs a perfect rotation around a center~$p$ (which can lie outside of~$Q$),
  scaling around~$p$ to remain in contact with the same edges of~$Q$.
\end{theorem}

On a single step of~$\silhouette$, the same three corner-edge
incidences between~$P$ and~$Q$ hold. This means that either all three
corners of~$P$ lie (in the relative interior) of the same three edges
of~$Q$, or one corner is fixed at a vertex of~$Q$, one other corner
moves in (the relative interior) of an edge of~$Q$, and the third
corner moves along a line in the interior of~$Q$.

When all three corners move on edges of~$Q$, then either all three
corners move counter-clockwise along $\partial Q$
(Figure~\ref{fig:inscribed_rotation_three_in}), or one moves clockwise
while the other two move counter-clockwise
(Figure~\ref{fig:inscribed_rotation_three_out}).  The two cases depend
on the center of perfect rotation in Theorem~\ref{thm:perfect} lying
inside~$Q$ or not.  We will say that a corner moves \emph{forward} if
it moves counter-clockwise or is stationary, \emph{backward} if it
moves clockwise.

We can therefore classify the steps of~$\silhouette$ into ten
different types: let $\TT[\sigma, \sigma']$
be the type of step $\PP[\sigma, \sigma']$.
Then, $\TT[\sigma, \sigma']$ is
\begin{denseitems}
\item type $f_{w}$ when all three corners of $\Ptheta$
lie on $\partial Q$ and move forward during the step;
\item type $a_{r}$ when corner $\atheta$ moves backward during the step;
\item type $a_{b}$ when corner $\atheta$ is at a vertex of~$Q$,
corner~$\btheta$ moves on an edge of~$Q$,
and corner~$\ctheta$ lies in the interior of~$Q$ during the step;
\item type $a_{c}$ when corner $\atheta$ is at a vertex of~$Q$,
corner~$\ctheta$ moves on an edge of~$Q$, and
corner~$\btheta$ lies in the interior of~$Q$ during the step;
\item and types~$b_{r}, b_{c}, b_{a}, c_{r}, c_{a}, c_{b}$ are defined
  analogously.
\end{denseitems}
We will see that the type of a step depends entirely on the angles
formed by the lines supporting the edges of~$Q$ involved in the step's
corner-edge incidences.

\medskip

We then study in detail what happens at a pivot, that is, when we move
from one step of the silhouette to the next.  At a pivot, we add an
additional corner-edge incidence to the three incidences from the
previous step, and then go on to the next step where one of the
original three incidences no longer holds.  This means that at a
pivot, we have four corner-edge incidences, so at least one corner
must coincide with a vertex of~$Q$.  We distinguish \emph{$2$-pivots},
where two corners lie at vertices of~$Q$ (the third corner is then
necessarily in the interior of~$Q$), and \emph{$3$-pivots}, where only
one corner lies at a vertex, while the other two corners lie in the
relative interior of edges of~$Q$.

\begin{figure}
  \centerline{\includegraphics[width=0.6\textwidth]{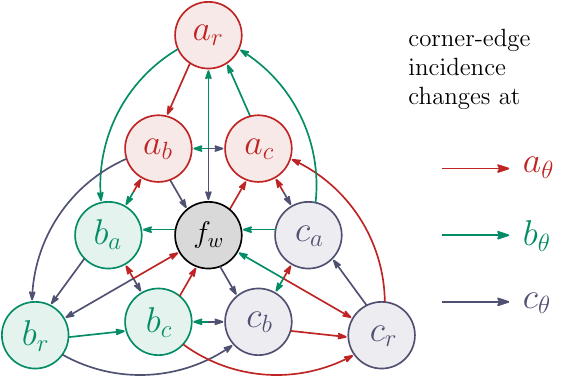}}
  \caption{Possible pairs of types, with arrows from~$e_1$ to~$e_2$.}
  \label{fig:state_diagram}
\end{figure}

In Section~\ref{sec:local_pivot} we study in detail how the type of
the step before and the step after the pivot relate. 
\begin{theorem}
  \label{thm:pivot}
  Let $P$ be a triangle and~$Q$ a convex polygon in general position,
  and consider the pivot~$\sigma$ between
  steps~$\PP[\sigma^{0}, \sigma]$ and~$\PP[\sigma, \sigma']$ of the
  silhouette~$\silhouette$.  Figure~\ref{fig:state_diagram} shows the
  pairs of types $\TT[\sigma^{0}, \sigma]$ and
  $\TT[\sigma, \sigma']$ that can occur, with the color of
  the arrow indicating which corner of $\PP(\sigma)$ changes its corner-edge incidence.
\end{theorem}

We next define a \emph{chain} to be a sequence of steps such that all
pivots of the chain are $3$-pivots and such that on all steps of the
chain the same two corners of~$P$ lie on the same two edges of~$Q$.
This immediately implies that these steps form a convex chain
of~$\silhouette$ inside a two-dimensional plane of~$\Reals^{4}$.

Our main theorem states that there is a decomposition of the
silhouette into a linear number of chains and a linear number of single steps.
\begin{theorem}
  \label{thm:main}
  Let $P$ be a triangle and~$Q$ a convex polygon in general position.
  Then the silhouette~$\silhouette$ of~$\CC$ can be decomposed
  into~$O(n)$ chains and single steps.  This subdivision can be
  computed in $O(n \log n)$ time.
\end{theorem}
To prove the Theorem~\ref{thm:main}, 
we define a \emph{stride} which consists of constant number of chains and steps,
and partition the global perfect rotation into linear number of strides
in Section~\ref{sec:final_partition}.
The proof relies heavily on the diagram of
Figure~\ref{fig:state_diagram}.

Since we obtain a linear number of strides that cover the global perfect rotation
sorted in counter-clockwise order with respect to the angle,
the chains and steps from Theorem~\ref{thm:main} are also
sorted in counter-clockwise order.
Then, for a given angle $\psi$, we can find
which chain or step contains $\PP(\psi)$ in $O(\log n)$ time using binary search,
in other words, we can find the three corner-edge incidences of $\PP(\psi)$.
Based on this fact, we compute the placement of $\PP(\psi)$ in $O(\log n)$ time.

Additionally, for each chain and step,
the number of pivots contained within them
can be computed in $O(\log n)$ time.
Thus, the complexity of silhouette $\Lambda$ can be computed
in $O(n \log n)$ time.
Since we will present that each step can be computed in $O(\log n)$ time,
we obtain an $O(k \log n)$-time algorithm to solve the problem $(2)$ and $(3)$
where $k$ is the complexity of silhouette $\Lambda$.

\subsection{Preliminaries}

Let $P = \triangle abc$ be a given triangle that contains the origin in its interior,
and let $Q$ be a given convex $n$-gon with no pair of parallel edges.
Suppose that orientation of three cornerss $a, b, c$ is counter-clockwise, and
we label the vertices of~$Q$ as~$q_{1}, q_{2},\ldots, q_{n}$
in counter-clockwise order, and the edges as~$e_{i} = q_{i}q_{i+1}$
for each $i \in [1, n]$.  For an integer $k \notin [1, n]$,
let $q_{k} = q_{r}$ if $k \equiv r \pmod{n}$ with $r \in [1, n]$.
For an integer pair $(i, j)$, let $q_{ij}$ be the intersection
point of the two lines $\overleftrightarrow{q_{i}q_{i+1}}$ and
$\overleftrightarrow{q_{j}q_{j+1}}$ (it exists since there are no
parallel edges),

Let $\partial Q$ be the boundary of $Q$ and
for any two distinct points $x, y \in \partial Q$,
let $\partial Q[x, y]$ denote the restricted boundary of $Q$
from $x$ to $y$ in the counter-clockwise direction, including both endpoints.
We naturally extend this notation to open and half-open intervals,
$\partial Q(x, y)$, $\partial Q[x, y)$, $\partial Q(x, y]$,
by excluding the corresponding endpoints.

For two points $u$ and $v$, let $|uv|$ denote
the euclidean distance of line segment $\overline{uv}$.

For two vectors $\vec{u}$, $\vec{v}$,
let $\phi(\vec{u}, \vec{v}) \in (-\pi, \pi]$ denote the counterclockwise angle
from vector $\vec{v}$ to vector $\vec{u}$.
For two integers $i, j$, let $\phi(i, j)  = \phi(\overrightarrow{q_{i+1}q_{i}},
 \overrightarrow{q_{j}q_{j+1}})$ and
for three points $x, y, z$ in the plane,
let $\angle xyz = \phi(\overrightarrow{yx}, \overrightarrow{yz})$.
For simplicity, let $\angle a, \angle b, \angle c$ be the angles
$\angle cab, \angle abc, \angle bca$, respecitvely,
and let $\angle q_{i} = \angle q_{i-1}q_{i}q_{i+1}$ for each $i \in [1, n]$.

For two triangles $\triangle uvw$ and $\triangle u'v'w'$,
if two triangles are similar each other,
we denote $\triangle uvw \sim \triangle u'v'w'$.
Let $[uvw]$ be the area of triangle $\triangle uvw$.

\section{The rotation center for a triangle inside a triangle}

Key to proving Theorem~\ref{thm:perfect} is an understanding of how a
triangle rotates inside another triangle.
\begin{definition}
  \label{def:center}
  The \emph{center of perfect rotation} of $\triangle xyz$ with
  respect to~$\triangle abc$ is the point $s$ which satisfies:
  \begin{enumerate}
  \item Let $h_{a}, h_{b}, h_{c}$ be the perpendicular feet from point~$s$
    to lines $\overleftrightarrow{yz}, \overleftrightarrow{zx},
    \overleftrightarrow{xy}$, respectively.
    Then, $|h_{a}h_{b}| : |h_{b}h_{c}| : |h_{c}h_{a}| = |ab| : |bc| : |ca|$.
  \item The point $s$ lies in the circumcircle of~$\triangle xyz$.
  \end{enumerate}
\end{definition}
The triangle obtained by projecting a point on the sides of a triangle
is called the \emph{pedal triangle}.  The definition states that~$s$
is the center of perfect rotation if its pedal triangle is similar
to~$\triangle abc$.

\begin{figure}
  \centerline{\includegraphics[width=0.5\textwidth]{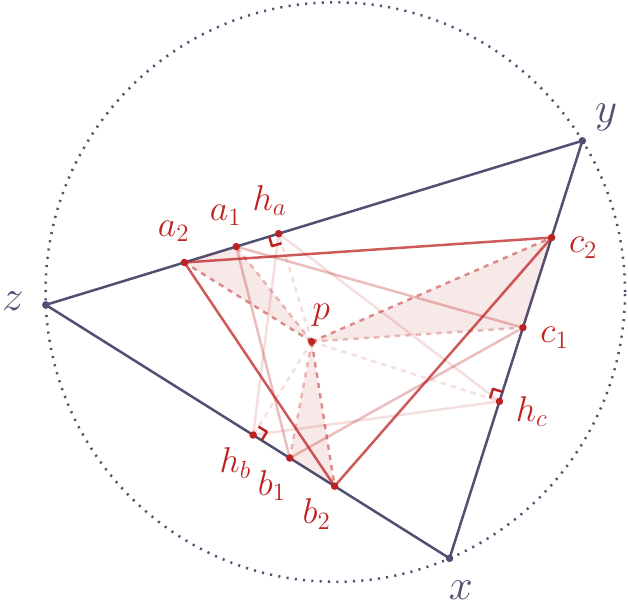}}
  \caption{The center of perfect rotation~$s$}
  \label{fig:center_of_inscribed_rotation}
\end{figure}

We call the point~$s$ the \emph{center of perfect rotation}
because of the following:
Let $\triangle a_{1}b_{1}c_{1}$ be an inscribed triangle of~$\triangle xyz$
that is similar to~$\triangle abc$ and
where the three vertices $a_{1}, b_{1}, c_{1}$ lie on the three edges
$\overline{yz}, \overline{zx}, \overline{xy}$, respectively.
Unless one vertex of $\triangle a_{1}b_{1}c_{1}$ coincides with
a vertex of $\triangle xyz$, we can slightly rotate $\triangle a_{1}b_{1}c_{1}$
around the center~$s$ and scale it from~$s$ until its vertices lie on
the three edges $\overline{yz}, \overline{zx}, \overline{xy}$ again.
This way we obtain another triangle $\triangle a_{2}b_{2}c_{2}$ that
is inscribed to $\triangle xyz$ and that is again similar
to~$\triangle abc$ (the three triangles $\triangle sa_{1}a_{2},
\triangle sb_{1}b_{2}, \triangle sc_{1}c_{2}$ are similar to each
other, see Figure~\ref{fig:center_of_inscribed_rotation}).

When $\triangle abc$ is an equilateral triangle, the point~$s$ is
a well-known triangle-center, the \emph{first isodynamic point},
the intersection of the three Apollonius circles of~$\triangle xyz$. 
For general~$\triangle abc$, the following lemma guarantees
the existence and uniqueness of the center of perfect rotation
and shows how to find it.
\begin{lemma}
  \label{lem:center}
  Given two triangles~$\triangle xyz$ and~$\triangle abc$, there
  exists a unique center of perfect rotation of~$\triangle xyz$ with
  respect to~$\triangle abc$.  This center can be computed in constant
  time.
\end{lemma}
\begin{proof}
  Let $\angle x, \angle y, \angle z$ be the angles
  $\angle zxy, \angle xyz, \angle yzx$, respectively.
  
  We first assume that all three $\angle a + \angle x, \angle b +
  \angle y, \angle c + \angle z$ are smaller than $\pi$.  Since
  $\angle a + \angle b + \angle c + \angle x + \angle y + \angle z =
  2\pi$, at least two of three angles $\angle a + \angle x, \angle b +
  \angle y, \angle c + \angle z$ are not smaller than $\pi/2$.
  Without loss of generality, let $\angle a + \angle x \geq \angle b +
  \angle y \geq \pi/2$.  Then, the set of points $u (\neq y, z)$ which
  satisfy $\angle zuy = \angle a + \angle x$ is an arc $\Gamma_{1}$
  with center $o_{1}$, and the set of points $u' (\neq z, x)$ which
  satisfy $\angle xu'z = \angle b + \angle y$ is an arc $\Gamma_{2}$
  with center $o_{2}$.

  We claim that two arcs $\Gamma_{1}$ and $\Gamma_{2}$ intersect
  only at point $s$ which is the center of perfect rotation of
  $\triangle xyz$ with respect to $\triangle abc$.  By the definitions
  of arc $\Gamma_{1}$ and center $o_{1}$, the angle $\angle o_{1}zy$
  can be computed as follows:
  \begin{align*}
    \angle o_{1}zy = \frac{1}{2} (\pi - \angle yo_{1}z) = \frac{1}{2} (\pi - 2(\pi - \angle zuy))
    = \angle zuy - \frac{\pi}{2} = \angle a + \angle x - \frac{\pi}{2}
  \end{align*}
  
  \noindent and, cyclically, we have $\angle xzo_{2} = \angle b + \angle y - \pi/2$.
  With these two angles, the angle $\angle o_{1}zo_{2}$ can be computed as follows:
  \begin{align*}
    \angle o_{1}zo_{2} = \angle o_{1}zy + \angle z + \angle xzo_{2}
    = (\angle a + \angle x - \frac{\pi}{2}) + \angle z + (\angle b + \angle y - \frac{\pi}{2})
    = \angle a + \angle b
  \end{align*}
  
  \noindent which is smaller than $\pi$.
  This implies that two arcs $\Gamma_{1}$ and $\Gamma_{2}$ intersect once at point $s$
  which is point $z$ mirrored by line $\overleftrightarrow{o_{1}o_{2}}$.
  We also observe that the angle $\angle ysx = 2\pi - \angle zsy - \angle xsz = \angle c + \angle z$
  is smaller than $\pi$, so that the point $s$ lies in the interior of $\triangle xyz$,
  which also means, lies in the circumcircle of $\triangle xyz$.
  
  Let $h_{a}, h_{b}, h_{c}$ be the perpendicular feet from point $s$
  to lines $\overleftrightarrow{yz}, \overleftrightarrow{zx}, \overleftrightarrow{xy}$.
  Since the four points $(s, h_{a}, z, h_{b})$ and four points $(s, h_{c}, y, h_{a})$
  are concyclic, we have $\angle sh_{a}h_{b} = \angle szh_{b}$ and 
  $\angle h_{c}h_{a}s = \angle h_{c}ys$.
  Then, the angle $\angle h_{c}h_{a}h_{b}$ can be computed as follows:
  \begin{align}
    \angle h_{c}h_{a}h_{b} = \angle h_{c}h_{a}s + \angle sh_{a}h_{b} 
    = \angle h_{c}ys + \angle szh_{b} 
    = \angle zsy - \angle zxy
    = \angle a
  \end{align}
  and cyclically, we have $\angle h_{a}h_{b}h_{c} = \angle b$
  and $\angle h_{b}h_{c}h_{a} = \angle c$.
  Therefore, $\triangle h_{a}h_{b}h_{c}$ is similar to $\triangle abc$ which implies that
  the point $s$ is the center of perfect rotation.
  
  We next assume that one of three $\angle a + \angle x, \angle b +
  \angle y, \angle c + \angle z$ is larger than $\pi$.  Since $\angle
  a + \angle b + \angle c + \angle x + \angle y + \angle z = 2\pi$, at
  least one of three $\angle a + \angle x, \angle b + \angle y, \angle
  c + \angle z$ is smaller than $\pi/2$.  Without loss of generality,
  let $\angle a + \angle x > \pi > \pi/2 > \angle c + \angle z$.
  Then, the set of points $v (\neq y, z)$ which satisfy $\angle yvz =
  2\pi - \angle a - \angle x$ is an arc $\Gamma_{3}$ with center
  $o_{3}$, and the set of points $v' (\neq z, x)$ which satisfy
  $\angle yv'x = \angle c + \angle z$ is an arc $\Gamma_{4}$ with
  center $o_{4}$.
 
  We again claim that two arcs $\Gamma_{3}$ and $\Gamma_{4}$ intersect
  only at point $s$.  By the
  definitions of two arcs $\Gamma_{3}, \Gamma_{4}$ and two centers
  $o_{3}, o_{4}$, we have $\angle o_{3}yz = 3\pi/2 - \angle a - \angle
  x$ and $\angle xyo_{4} = \pi/2 - \angle c - \angle z$.  Since
  $\angle o_{3}yz + \angle xyo_{4} = 2\pi - \angle a - \angle x -
  \angle c - \angle z = \angle b + \angle y$ is larger than $\angle
  y$, we also obtain $\angle xyo_{4} > \angle xyo_{3}$, in other
  words, $\angle o_{3}yo_{4} > 0$.  And the circle containing the arc
  $\Gamma_{3}$ passes through the vertex $z$, whereas the disk
  containing the arc $\Gamma_{4}$ on its boundary does not contain the
  vertex $z$ because $2 \cdot \angle z < \angle yo_{4}x$.  Thus, by
  these two observations, two arcs $\Gamma_{3}$ and $\Gamma_{4}$ must
  intersect only at point $s$ which lies in the exterior of $\triangle
  xyz$.
 
  Since the sum of two angles $\angle ysz + \angle x = 2\pi - \angle a
  - \angle x + \angle x = 2\pi - \angle a$ is larger than $\pi$, the
  point $s$ lies in the circumcircle of $\triangle xyz$.  And, by
  applying the equation $(1)$ cyclically, we have $\angle
  h_{a}h_{b}h_{c} = \angle b$ and $\angle h_{c}h_{a}h_{b} = \angle c$.
  Therefore, $\triangle h_{a}h_{b}h_{c}$ is similar to $\triangle abc$
  which implies the point $s$ is the center of perfect rotation.
 
  The remaining case is one of three $\angle a + \angle x, \angle b +
  \angle y, \angle c + \angle z$ equals $\pi$; without loss of
  generality, let $\angle a + \angle x = \pi$.  Then, we always can
  find the point $s$ on the line segment $\overline{yz}$ such that
  $\angle xsz = \angle b + \angle y = \pi - \angle c - \angle z = \pi
  - \angle ysx$.  Let $h_{b}, h_{c}$ be the perpendicular feet from
  point $s$ to lines $\overleftrightarrow{zx},
  \overleftrightarrow{xy}$.  By the concyclicity of four points $(s,
  h_{b}, x, h_{c})$, we have $\angle sh_{b}h_{c} = \angle sxh_{c} =
  \angle xsz - \angle y = \angle b$ and $\angle h_{b}h_{c}s = \angle
  h_{b}xs = \angle ysx - \angle z = \angle c$.  Thus, if we consider
  the point $s$ as the perpendicular feet $h_{a}$ from $s$ to line
  $\overleftrightarrow{yz}$, then $\triangle h_{a}h_{b}h_{c}$ is
  similar to $\triangle abc$.  Since the point $s$ also lies in the
  circumcircle of $\triangle xyz$ (lies on $\overline{yz}$), the point
  $s$ is the center of perfect rotation.
 
  So far, we prove that the center of perfect rotation of $\triangle
  xyz$ with respect to $\triangle abc$ always exists and also show
  that how to compute it with two arcs in constant time.  We will
  finish the proof by showing that the center of perfect rotation
  $s$ is unique.
 
  Let $\Gamma_{a}$ be the circumcircle of the four points $(s, h_{b},
  x, h_{c})$ and define the two circumcircles $\Gamma_{b}$ and
  $\Gamma_{c}$, cyclically.  Since the line segments $\overline{sx},
  \overline{sy}, \overline{sz}$ are the diameters of circles
  $\Gamma_{a}, \Gamma_{b}, \Gamma_{c}$, respectively, by the sine law,
  \begin{align}
    |sx| : |sy| : |sz| = \frac{|h_{b}h_{c}|}{\sin \angle x} :
    \frac{|h_{c}h_{a}|}{\sin \angle y}
    : \frac{|h_{a}h_{b}|}{\sin \angle z}
    = \frac{|bc|}{\sin \angle x} : \frac{|ca|}{\sin \angle y} :
    \frac{|ab|}{\sin \angle z}
  \end{align}
  The set of points~$w$ which satisfy $|wy| : |wz| = |ca|/\sin \angle
  y : |ab|/\sin \angle z$ is a circle $\Gamma_{yz}$ and cyclically, we
  obtain the two circles $\Gamma_{zx}$ and $\Gamma_{xy}$.  There are
  two intersections~$s$ and~$s'$ of the three circles $\Gamma_{yz},
  \Gamma_{zx}, \Gamma_{xy}$, which are inverse to each other with
  respect to the circumcircle of~$\triangle xyz$
  (the proof is analogous to establishing the inverse relationship
  between the first and second isodynamic points with respect to the circumcircle).
  Therefore, the center of perfect rotation is unique.
\end{proof}

Lemma~\ref{lem:center} also implies that the point $s$ is the center
of perfect rotation of $\triangle xyz$ with repsect to~$\triangle  abc$
if and only if $\angle zsy = \angle a + \angle x, \angle xsz =
\angle b + \angle y, \angle ysx = \angle c + \angle z$.
Based on this fact, we prove the following lemma.

\begin{lemma}[Trilinear Coordinates]
  \label{lem:tri_coordinates}
  For a center of perfect rotation $s$ of $\triangle xyz$ with respect to $\triangle abc$,
  let $d(a)$ be the directed distance from $h_{a}$ to $s$
  such that $d(a) = |h_{a}s|$ when $s$ lies on the same side of line $\overleftrightarrow{yz}$
  from the interior of $\triangle xyz$
  and $d(a) = -|h_{a}s|$ when $s$ lies on the opposite side of line $\overleftrightarrow{yz}$
  from the interior of $\triangle xyz$,
  and we define $d(b), d(c)$, cyclically.
  Then,
  \begin{align*}
    d(a) : d(b) : d(c) =
    \frac{\sin (\angle a + \angle x)}{\sin \angle a} :
    \frac{\sin(\angle b + \angle y)}{\sin \angle b} :
    \frac{\sin(\angle c + \angle z)}{\sin \angle c}.
  \end{align*}
\end{lemma}
\begin{proof}
 By the definition, we can compute the area of $\triangle syz$
 in two different ways as follows:
 \begin{align*}
  [syz] = \frac{1}{2} \cdot |sy| \cdot |sz| \cdot |\sin(\angle zsy)|
  = \frac{1}{2} \cdot |sh_{a}| \cdot |yz|
 \end{align*}
 Since the center $s$ lies on the opposite side of line $\overleftrightarrow{yz}$
 from the interior of $\triangle xyz$ when $\angle zsy > \pi$,
 the directed distance $d(a)$ can be expressed as:
 \begin{align*}
  d(a) = \frac{|sy| \cdot |sz| \cdot \sin(\angle a + \angle x)}{|yz|}
 \end{align*}
 Then, by the equation $(2)$ from the proof of Lemma~\ref{lem:center} and the sine law,
 the ratio
  \begin{align*}
  d(a) : d(b) : d(c) &= \frac{|sy| \cdot |sz| \cdot \sin (\angle a + \angle x)}{|yz|}
  : \frac{|sz| \cdot |sx| \cdot \sin (\angle b + \angle y)}{|zx|}
  : \frac{|sx| \cdot |sy| \cdot \sin (\angle c + \angle z)}{|xy|} \\
  &= \frac{\sin (\angle a + \angle x)}{|sx| \cdot |yz|}
  : \frac{\sin(\angle b + \angle y)}{|sy| \cdot |zx|}
  : \frac{\sin(\angle c + \angle z)}{|sz| \cdot |xy|} \\
  &= \frac{\sin \angle x \cdot \sin (\angle a + \angle x) }{|bc| \cdot |yz|}
  : \frac{\sin \angle y \cdot \sin (\angle b + \angle y)}{|ca| \cdot |zx|}
  : \frac{\sin \angle z \cdot \sin (\angle c + \angle z) }{|ab| \cdot |xy|} \\
  &= \frac{\sin (\angle a + \angle x)}{\sin \angle a}
  : \frac{\sin (\angle b + \angle y)}{\sin \angle b}
  : \frac{\sin (\angle c + \angle z)}{\sin \angle c},
 \end{align*}
 proving the lemma.
\end{proof}

In the following section, we will prove
Theorem~\ref{thm:perfect} with Lemma~\ref{lem:center}
and present how to decide the type of step with Lemma~\ref{lem:tri_coordinates}.
 
\section{Perfect rotation along a step of $\Lambda$}
\label{sec:local}

By Theorem~\ref{thm:continuous}, the three vertices $\atheta, \btheta, \ctheta$
move continuously during the global perfect rotation~$\PP[0, 2\pi]$.
For every~$\theta$, we define the \emph{triplet}
$\Etheta = (a(\theta), b(\theta), c(\theta))$ as follows:
$a(\theta)$ is the index of the edge of~$Q$ that~$\atheta$ lies on.
If corner $\atheta$ coincides with vertex $q_{k} = e_{k-1} \cap e_{k}$ of~$Q$,
then~$a(\theta) = k$ (rather than~$k-1$).
If~$\atheta$ lies in the interior of~$Q$,
then $a(\theta)$ is the index of the last edge that corner $\atheta$ lies on,
in other words, the edge that corner $\atheta$ departs from $\partial Q$.
If corner $\atheta$ moves forward and passes through the vertex $q_{1}$,
we will say, $a(\theta) = n+1$ rather than $1$.
We define $b(\theta)$ and $c(\theta)$ analogously.

The triplet~$\Etheta$ remains constant
along a step of the silhouette~$\silhouette$,
since the halfspaces defining the containment polytope~$\CC$
are exactly the incidences between corners of~$P$ and edges of~$Q$.
Thus, for a given pivot~$\sigma$ and the next pivot~$\sigma'$ of~$\sigma$,
the triple~$\EE(\sigma')$ can differ
from~$\EE(\sigma)$ in only one position.
With this definition of $\Etheta$, we prove Theorem~\ref{thm:perfect}.

\begin{proof}[Proof of Theorem~\ref{thm:perfect}.]
We observe a perfect rotation $\PP(\sigma, \sigma')$
from a step $\PP[\sigma, \sigma']$
and by the definition, there is no pivot in the interval $(\sigma, \sigma')$.

We first consider the case when a corner of~$\Ptheta$ lies
in the interior of~$Q$, for instance,
corner $\btheta$ lies in the interior of $Q$ during the step.
Since there are three vertex-edge incidences 
during the perfect rotation $\PP(\sigma, \sigma')$,
we have $\atheta = q_{i}$ or~$\ctheta = q_{k}$.

Without loss of generality we assume
$\atheta = q_{i}$, which implies that the angles~$\phi(i, k)$ and
$\phi(k, i-1)$ are positive (see Figure~\ref{fig:inscribed_rotation_two}).
Under these assumptions, as~$\Ptheta$ rotates,
the corner~$\atheta$ remains fixed at $q_{i}$,
the corner~$\ctheta$ moves along the edge~$e_{k}$
in counter-clockwise direction and
the corner~$\btheta$ moves in the interior of~$Q$.
Symmetrically, if the corner $\ctheta$ lies in the interior of $Q$,
the corner~$\btheta$ moves along the edge~$e_{j}$ in counter-clockwise direction
and the corner~$\ctheta$ moves inside the interior of~$Q$.
Therefore, if a vertex of $\Ptheta$
lies in the interior of~$Q$ during the perfect rotation,
one vertex of $\Ptheta$, which is the center $p$,
remains fixed at a vertex of~$Q$
and the other vertex of~$\Ptheta$ moves
in counter-clockwise direction along $\partial Q$.

\begin{figure}
  \centerline{\includegraphics[width=1.1\textwidth]{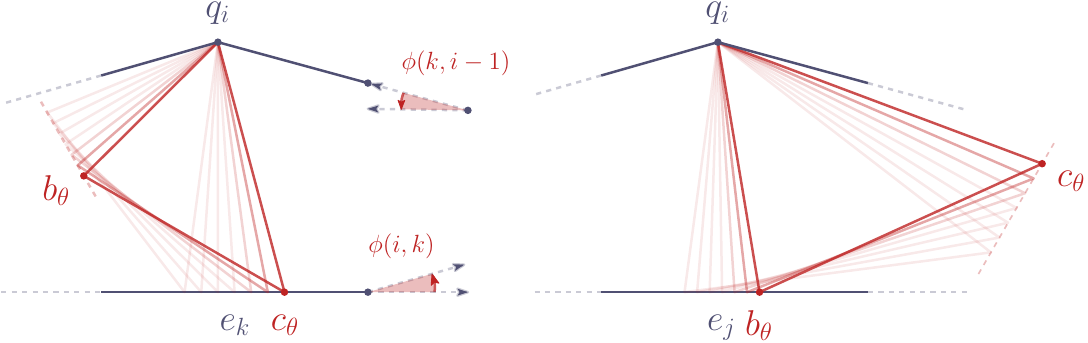}}
  \caption{Perfect rotations when a vertex of $\Ptheta$ lies in the interior of $Q$.}
  \label{fig:inscribed_rotation_two}
\end{figure}

We now assume that all three corners of $\Ptheta$ lie on $\partial Q$
and $\Etheta = (i, j, k)$ for $\theta \in (\sigma, \sigma')$.
Then, we construct~$\triangle q_{jk}q_{ki}q_{ij}$ by extending
the three edges~$e_{i}, e_{j}, e_{k}$ and find the center of rotation~$s_{ijk}$
of $\triangle q_{jk}q_{ki}q_{ij}$ with respect to $P = \triangle abc$
(see Figure~\ref{fig:inscribed_rotation_three_in}
and Figure~\ref{fig:inscribed_rotation_three_out}).
By Lemma~\ref{lem:center}, $\Ptheta$ rotates during the step
with the center of perfect rotation $p = s_{ijk}$.
\end{proof}

We define the \emph{moving distance} $d(\atheta, \delta)$
as the distance from~$\atheta$ to~$a_{\theta + \delta}$
along~$\partial Q$ in counter-clockwise direction.
By definition, the moving distance is negative when~$\atheta$ moves
backwards. We make analogous definitions for~$\btheta$ and~$\ctheta$,
and have the following lemma.

\begin{corollary}
\label{cor:moving_dists}
 If $\Ttheta = \TT(\theta + \delta) = (i, j, k)$ for some angle $\delta$,
 \begin{align*}
  d(\atheta, \delta) : d(\btheta, \delta) : d(\ctheta, \delta)
  = \frac{\sin (\angle a + \phi(j, k))}{\sin \angle a}
  : \frac{\sin (\angle b + \phi(k, i))}{\sin \angle b}
  : \frac{\sin (\angle c + \phi(i, j))}{\sin \angle c}.
 \end{align*}
 In particular, for a given $3$-pivot $\sigma$,
 we can compute the next pivot $\sigma'$ of $\sigma$ in constant time.
\end{corollary}
\begin{proof}
 For $\Ptheta = \triangle \atheta\btheta\ctheta$ and $\PP(\theta +
 \delta) = \triangle a_{\theta + \delta}b_{\theta + \delta}c_{\theta +
   \delta}$ with $\Ttheta = \TT(\theta + \delta) = (i, j, k)$, let $p$
 be the center $s_{ijk}$.  If $h_{a}, h_{b}, h_{c}$ are perpendicular
 feet from $p$ to lines $\overleftrightarrow{e_{i}},
 \overleftrightarrow{e_{j}}, \overleftrightarrow{e_{k}}$,
 respectively, all $\triangle sh_{a}\atheta, \triangle sh_{b}\btheta,
 \triangle sh_{c}\ctheta$ are similar to each other and all $\triangle
 sh_{a}a_{\theta + \delta}$, $\triangle sh_{b}b_{\theta + \delta}$,
 $\triangle sh_{c}c_{\theta + \delta}$ are similar to each
 other. Then, all $\triangle s\atheta a_{\theta + \delta}$, $\triangle
 s\btheta b_{\theta + \delta}$, $\triangle s\ctheta c_{\theta +
   \delta}$ are similar to each other with scaling ratio 
   $d(a) : d(b) : d(c)$ from Lemma~\ref{lem:tri_coordinates}.
\end{proof}

To denote the angles formed by the lines supporting the edges of~$Q$
containing the vertices of~$\Ptheta$,
we introduce~$\phi_{ab}(\theta) = \phi(a(\theta), b(\theta))$,
with~$\phi_{bc}(\theta)$ and $\phi_{ca}(\theta)$ defined cyclically.

Corollary~\ref{cor:moving_dists} implies that if all three angles
$\angle a + \phi_{bc}(\theta), \angle b + \phi_{ca}(\theta),
\angle c + \phi(i, j)$ are smaller than~$\pi$,
then all corners $\atheta, \btheta, \ctheta$ of $\Ptheta$ move forward,
as the center of inscribed rotation~$s_{ijk}$ lies
in the interior of $\triangle q_{jk}q_{ki}q_{ij}$.

If one of the three angles equals~$\pi$, let's say,
$\angle a + \phi_{bc}(\theta) = \pi$, then corner $\atheta$ remains fixed
and two corners $\btheta, \ctheta$ move forward
(the center~$s_{ijk}$ coincides with~corner $\atheta$).
If one of the three angles is larger than~$\pi$,
let's say, $\angle a + \phi_{bc}(\theta) > \pi$,
then corner $\atheta$ moves backward
(the center~$s_{ijk}$ lies outside of~$\triangle q_{jk}q_{ki}q_{ij}$).
Since $\angle a + \angle b + \angle c + \phi_{bc}(\theta)
+ \phi_{ca}(\theta) + \phi_{ab}(\theta) = 2\pi$,
only one corner of $\Ptheta$ can be fixed or move backward.

Therefore, if all three corners of $\Ptheta$ lie on $\partial Q$ 
then the moving directions and relative speeds
of~$\atheta, \btheta, \ctheta$ are determined \emph{only} by
the three angles~$\phi_{bc}(\theta)$, $\phi_{ca}(\theta)$
and~$\phi_{ab}(\theta)$.
In other words, the moving directions and relative speeds of
three corners~$\atheta$, $\btheta$, $\ctheta$ of $\Ptheta$
are invariant in the interior of every step of~$\silhouette$.

In summary, for a given step $\PP[\sigma, \sigma']$,
we can decide the type of step $\TT[\sigma, \sigma']$
by computing three angles $\angle a + \phi_{bc}(\theta)$,
$\angle b + \phi_{ca}(\theta)$, $\angle c + \phi_{ab}(\theta)$ as follows:
for $\theta \in (\sigma, \sigma')$,
\begin{enumerate}
 \item $\TT[\sigma, \sigma'] = f_{w}$:
 all three angles $\angle a + \phi_{bc}(\theta)$,
 $\angle b + \phi_{ca}(\theta)$, $\angle c + \phi_{ab}(\theta)$
 are not larger than $\pi$ .
 \item $\TT[\sigma, \sigma'] = a_{r}$:
 when $\angle a + \phi_{bc}(\theta) > \pi$.
 \item $\TT[\sigma, \sigma'] = a_{c}$:
 when $\phi_{ca}(\theta) < 0 < \phi(c(\theta), a(\theta)-1)$.
 \item $\TT[\sigma, \sigma'] = a_{b}$:
 when $\phi(a(\theta)-1, b(\theta)) < 0 < \phi_{ab}(\theta)$.
\end{enumerate}
and types $b_{r}$, $b_{c}$, $b_{a}$, $c_{r}$, $c_{a}$, $c_{b}$
can be found analogously. 
Based on this fact, we prove the following corollary.
\begin{corollary}
\label{cor:two_triangles}
 For two angles $\alpha$ and $\beta$ such that
 $\alpha < \beta$ and four corners
 $b_{\alpha}$, $b_{\beta}$, $c_{\alpha}$, $c_{\beta}$ appears
 in counter-clockwise order along $\partial Q$.
 Then, if two corners $a_{\alpha}$, $b_{\beta}$ appears
 in counter-clockwise ($resp.$ clockwise) order along $\partial Q$,
 then the angle $\angle a + \phi(\overrightarrow{b_{\beta}b_{\alpha}},
 \overrightarrow{c_{\alpha}c_{\beta}})$ is not larger
 ($resp.$ larger) than $\pi$.
\end{corollary}
\begin{proof}
 Construct the triangle by extending three lines
 $\overleftrightarrow{a_{\alpha}a_{\beta}},
 \overleftrightarrow{b_{\alpha}b_{\beta}},
 \overleftrightarrow{c_{\alpha}c_{\beta}}$.
 Since $\alpha < \beta$, corner $a_{\alpha}$ moves forward or backward
 along the boundary of this triangle,
 depending on the order of $a_{\alpha}$ and $a_{\beta}$.
 Thus, we can decide whether $\angle a + \phi(\overrightarrow{b_{\beta}b_{\alpha}},
 \overrightarrow{c_{\alpha}c_{\beta}})$ is larger than $\pi$ or not.
\end{proof}

\section{What happens at a pivot of~$\silhouette$}
\label{sec:local_pivot}

\begin{figure}
  \centerline{\includegraphics[width=0.9\textwidth]{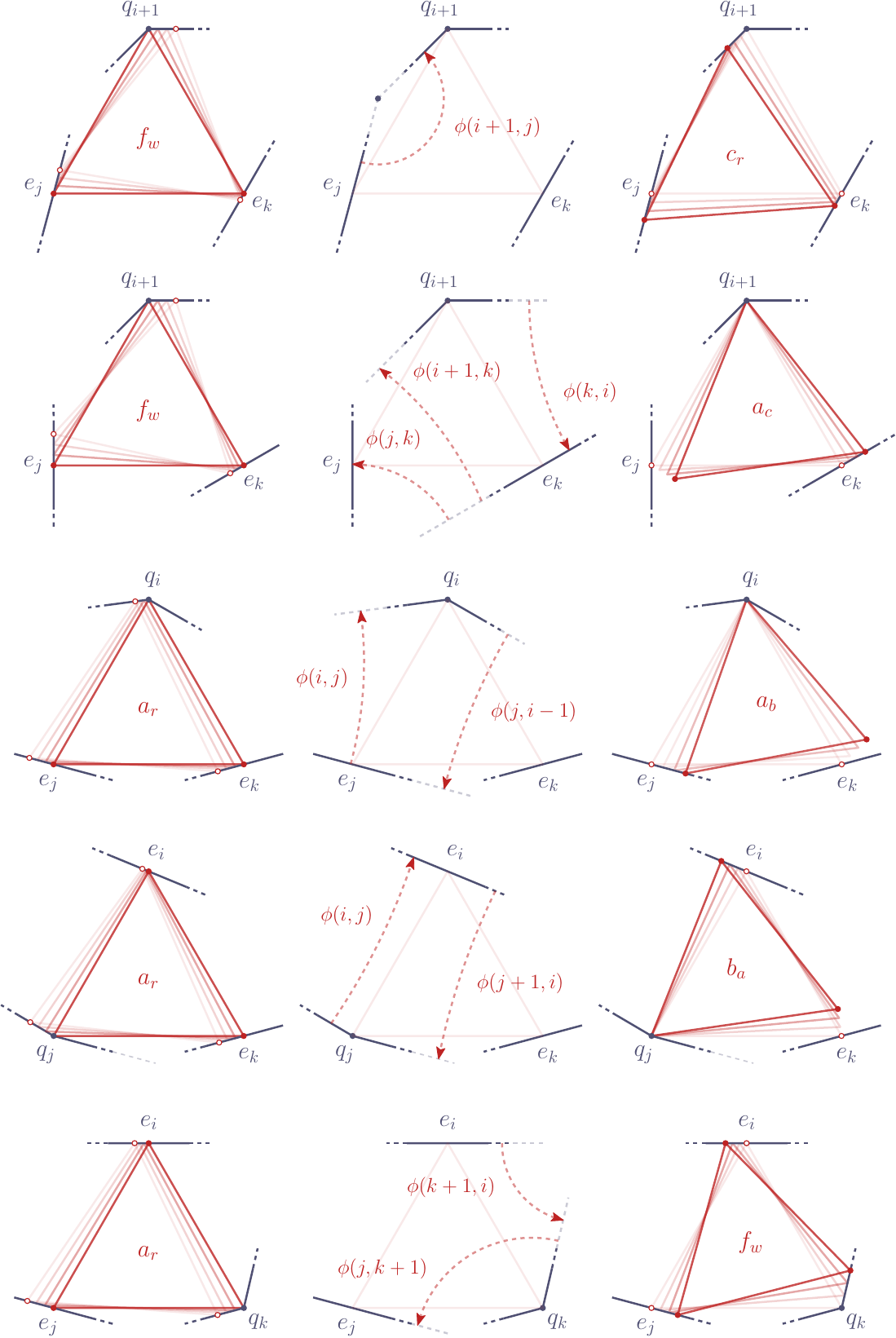}}
  \caption{Instances of $P$ and $Q$ for pair $(s_{1}, s_{2}) = (f_{w}, c_{r}), (f_{w}, a_{c}),
  (a_{r}, a_{b}), (a_{r}, b_{a}), (a_{r}, f_{w})$}
  \label{fig:pivot_first_five}
\end{figure}

\begin{figure}
  \centerline{\includegraphics[width=0.9\textwidth]{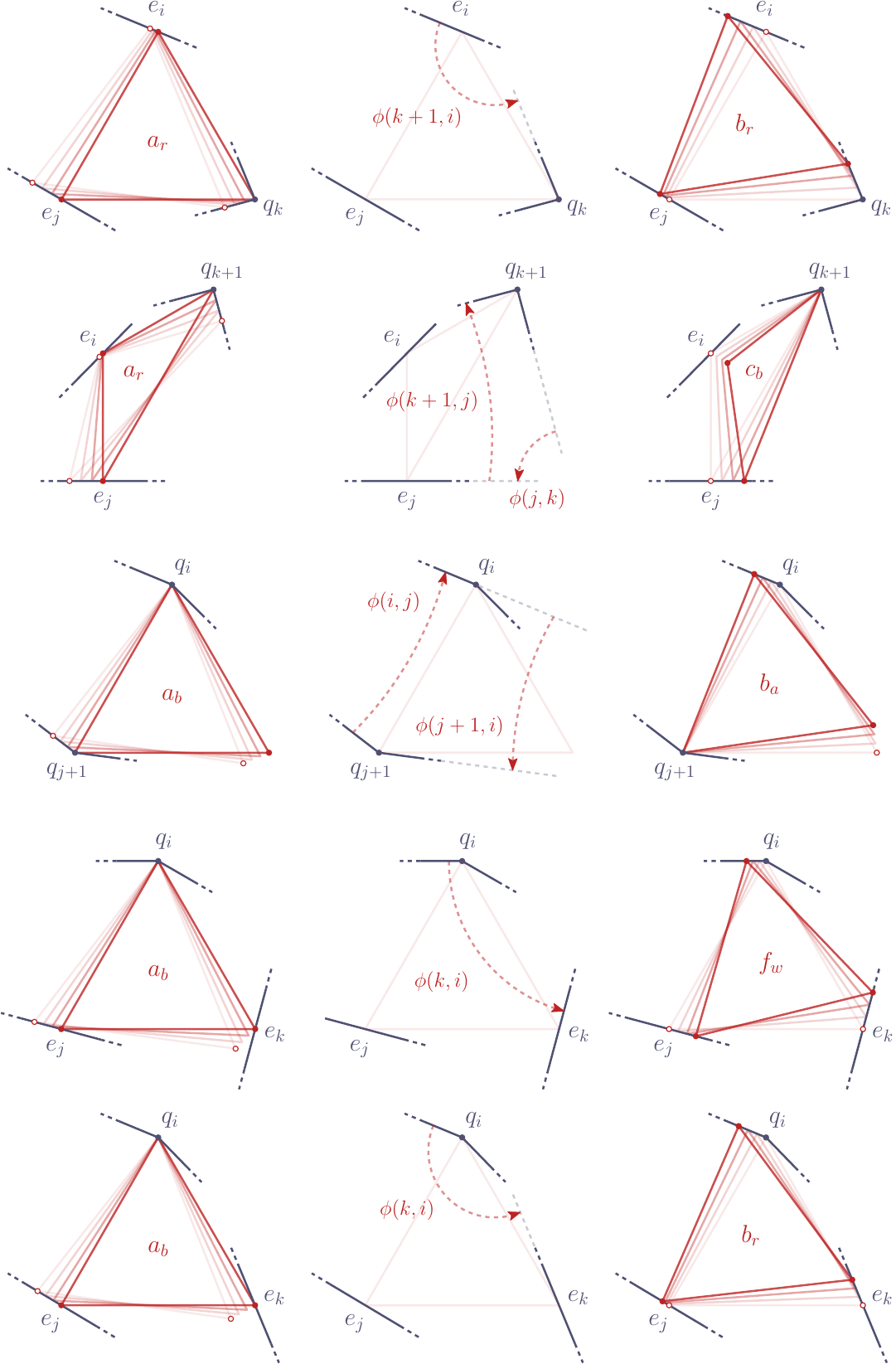}}
  \caption{Instances of $P$ and $Q$ for pair $(s_{1}, s_{2}) = (a_{r}, b_{r}), (a_{r}, c_{b}),
  (a_{b}, b_{a}), (a_{b}, f_{w}), (a_{b}, b_{r})$}
  \label{fig:pivot_second_five}
\end{figure}

\begin{figure}
  \centerline{\includegraphics[width=0.9\textwidth]{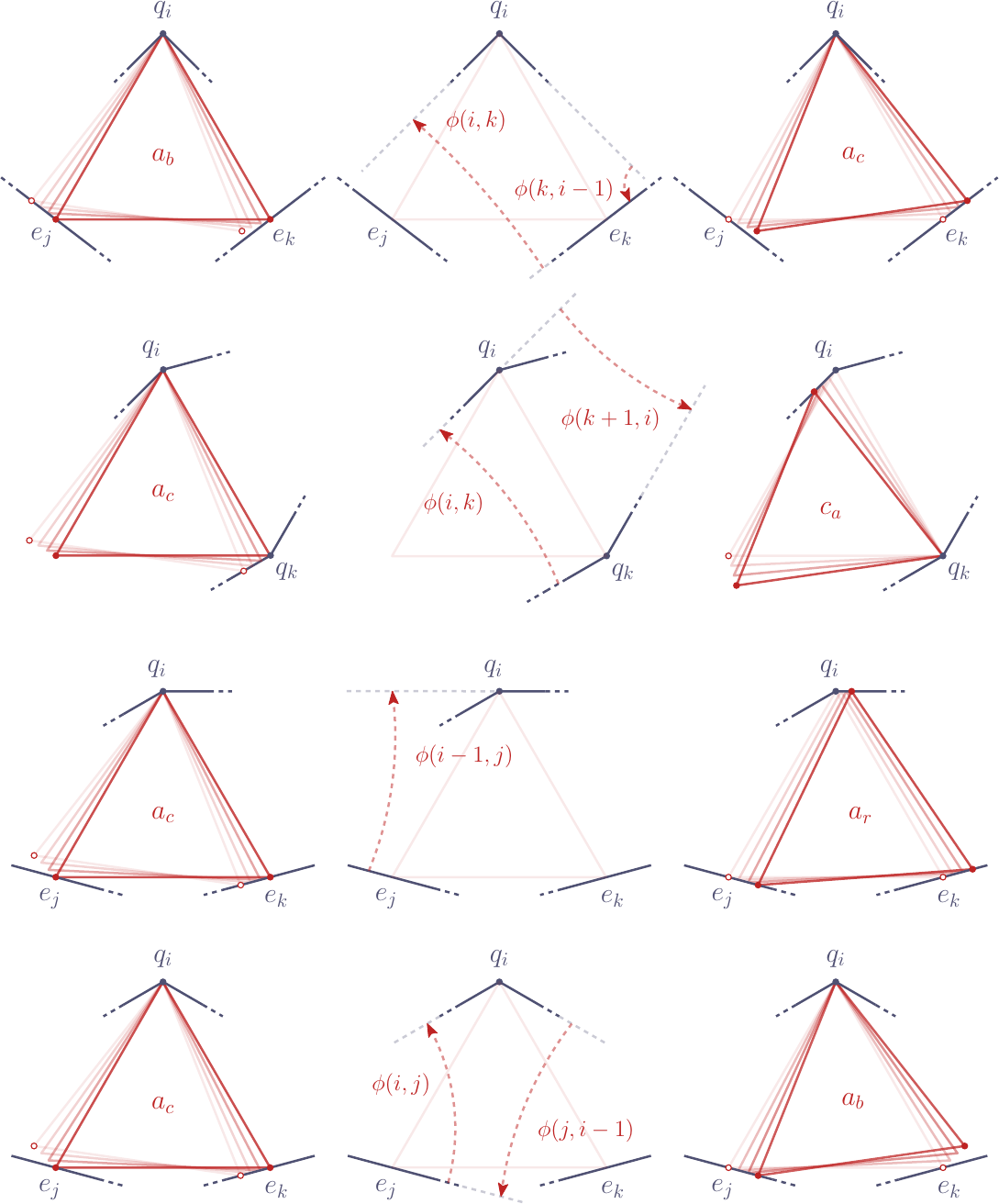}}
  \caption{Instances of $P$ and $Q$ for pair $(s_{1}, s_{2}) = (a_{b}, a_{c}), (a_{c}, c_{a}),
  (a_{c}, a_{r}), (a_{c}, a_{b})$}
  \label{fig:pivot_third_four}
\end{figure}

Consider a perfect rotation~$\PP[\theta_{1}, \theta_{2}]$ from a
point on one step of~$\silhouette$ to a point on the next step.
In other words, there is exactly one pivot~$\sigma$
with~$\theta_1 < \sigma < \theta_2$.  
Suppose that the type of step changes from $t_{1}$ to $t_{2}$ at pivot $\sigma$.
Then, by discussing the possible pairs of states~$(t_{1}, t_{2})$ that can occur,
we prove the Theorem~\ref{thm:pivot}
(see Figure~\ref{fig:pivot_first_five}, Figure~\ref{fig:pivot_second_five},
and Figure~\ref{fig:pivot_third_four} at the end of paper).

\begin{proof}[Proof of Theorem~\ref{thm:pivot}.]
Suppose that $\EE(\theta_{1}) = (i, j, k)$. Then we enumerate
all possible types $t_{2}$ depending on the type $t_{1}$.

\begin{itemize}
 \item When~$t_{1} = f_{w}$, without loss of generality,
 assume that $a_{\sigma}$ coincides with~$q_{i+1}$.
 Then~$\phi(i, j) < \phi(i+1, j)$ and $\phi(k, i) > \phi(k, i+1)$.
 \begin{itemize}
  \item If $\phi(i+1, j) > \pi - \angle c$, then~$t_{2} = c_{r}$
  since $c_{\sigma}$ moves backward.
  \item If $\phi(k, i+1) < 0$, then~$t_{2} = a_{c}$
  since $\angle a + \phi(j, k) < \pi$.
 \end{itemize}
 Thus, we have two possible pairs $(t_{1}, t_{2}) = (f_{w}, c_{r})$ or
 $(f_{w}, a_{c})$, the pairs~$(f_{w}, a_{r})$, $(f_{w}, b_{r})$,
 $(f_{w}, b_{a})$, $(f_{w}, c_{b})$ for the symmetric cases, and the
 trivial pair~$(f_{w}, f_{w})$.

 \bigskip
 \item When~$t_{1} = a_{r}$,
 \begin{itemize}
  \item if $a_{\sigma}$ coincides with $q_{i}$ and $\phi(i - 1, j) < 0$,
  then~$t_{2} = a_{b}$ since $\angle a + \phi(j, k) > \pi$.
  \item if $b_{\sigma}$ coincides with $q_{j+1}$ and $\phi(i, j+1) < 0$,
  then~$t_{2} = b_{a}$ since $\angle b + \phi(k, i) < \pi$.
  \item if $c_{\sigma}$ coincides with the vertex $q_{k+1}$ of $Q$ and
  \begin{itemize}
   \item if $\phi(j, k+1) \leq \pi - \angle a$,
   then~$t_{2} = f_{w}$ since $a_{\sigma}$ moves forward.
   \item if $\phi(k+1, i) > \pi - \angle b$, then~$t_{2} = b_{r}$
   since $b_{\sigma}$ moves backward.
   \item if $\phi(j, k+1) < 0$, then~$t_{2}= c_{b}$
   since $\angle c + \phi(i, j) < \pi$.
  \end{itemize}
 \end{itemize}
 Here, we have six possible pairs $(a_{r}, a_{b})$, $(a_{r}, b_{a})$,
 $(a_{r}, f_{w})$, $(a_{r}, b_{r})$, $(a_{r}, c_{b})$ and $(a_{r},
 a_{r})$.
 
 \bigskip
 \item When~$t_{1} = a_{b}$,
 \begin{itemize}
  \item if $b_{\sigma}$ coincides with $q_{j+1}$ and $\phi(i, j+1) < 0 < \phi(i, j)$,
    then~$t_{2} = b_{a}$.
  \item if $c_{\sigma}$ reaches on the edge $e_{k}$,
  then $\angle a + \phi(j, k) < \pi$. And
  \begin{itemize}
   \item if $\phi(k, i) \leq \pi - \angle b$, then~$t_{2} = f_{w}$
   since $b_{\sigma}$ moves forward.
   \item if $\phi(k, i) > \pi - \angle b$, then~$t_{2} = b_{r}$
   since $b_{\sigma}$ moves backward.
   \item if $\phi(k, i) < 0 < \phi(k, i-1)$, then~$t_{2} = a_{c}$
   since $\angle a + \phi(j, k) < \pi$.
  \end{itemize}
 \end{itemize}
 We have five possible pairs $(a_{b}, b_{a})$, $(a_{b}, f_{w})$,
 $(a_{b}, b_{r})$, $(a_{b}, a_{c})$ and $(a_{b}, a_{b})$.
 
 \bigskip
 \item When~$t_{1} = a_{c}$,
 \begin{itemize}
  \item if $c_{\sigma}$ coincides with $q_{k+1}$ and $\phi(i, k+1) < 0 < \phi(i, k)$,
  then~$t_{2} = c_{a}$.
  \item if $b_{\sigma}$ reaches on the edge $e_{j}$,
  then $\angle a + \phi(j, k) > \pi$.
  \begin{itemize}
   \item if $\phi(i-1, j) > 0$, then~$t_{2} = a_{r}$
   since $a_{\sigma}$ moves backward.
   \item if $\phi(i-1, j) < 0$, then~$t_{2} = a_{b}$
   since $\angle a + \phi(j, k) > \pi$.
  \end{itemize}
 \end{itemize}
 We have four possible pairs $(a_{c}, c_{a})$, $(a_{c}, a_{r})$,
 $(a_{c}, a_{b})$ and $(a_{c}, a_{c})$.
 \end{itemize}

\noindent The possible pairs~$(t_1, t_2)$ where $t_{1}$ is one of the
six types $b_{r}, c_{r}, b_{c}, c_{a}, b_{a}, c_{b}$ follow by
symmetry.  Figure~\ref{fig:state_diagram} shows the state diagram, with
all possible transitions indicated by arrows.
Additionally, two pairs $(a_{r}, b_{r}), (a_{r}, c_{b})$ can be obtained by
following $a_{r} \rightarrow f_{w} \rightarrow b_{r}$ and $a_{r}
\rightarrow f_{w} \rightarrow c_{b}$.
\end{proof}

To summarize, when~$\Ptheta$ arrives at a pivot~$\sigma$, one of the
three angles~$\phi_{ab}(\theta), \phi_{bc}(\theta), \phi_{ca}(\theta)$
increases, another one decreases, and the last one remains unchanged.
If one of the three angles $\phi_{ab}(\theta), \phi_{bc}(\theta),
\phi_{ca}(\theta)$ becomes negative, the type of step becomes one of
the six types $a_{b}, a_{c}, b_{a}, b_{c}, c_{a}, c_{b}$. 
If one of the three angles $\angle a + \phi_{bc}(\theta), \angle b +
\phi_{ca}(\theta),\angle c + \phi_{ab}(\theta)$ crosses the
value~$\pi$, then the type of step can become one of the four types $f_{w},
a_{r}, b_{r}, c_{r}$.

\bigskip
We next observe that some specific pivots restrict the placement of $\Ptheta$
as stated in the following Corollary~\ref{cor:placement_ac}
and Lemma~\ref{lem:placement_ar}.

\begin{corollary}
 \label{cor:placement_ac}
 For two consecutive pivots $\sigma$ and $\sigma'$ where
 the type of the step $\PP[\sigma, \sigma']$ is $a_{c}$ or $c_{a}$,
 there exists no orientation~$\theta \in [0, 2\pi]$ such that
 the four corners $a_{\sigma'}$, $\atheta$, $\ctheta$, and $c_{\sigma'}$
 appear in a counterclockwise order along $\partial Q$
 with $\atheta, \ctheta \in \partial Q(a_{\sigma'}, c_{\sigma'})$.
\end{corollary}

\begin{figure}
  \centerline{\includegraphics[width=0.8\textwidth]{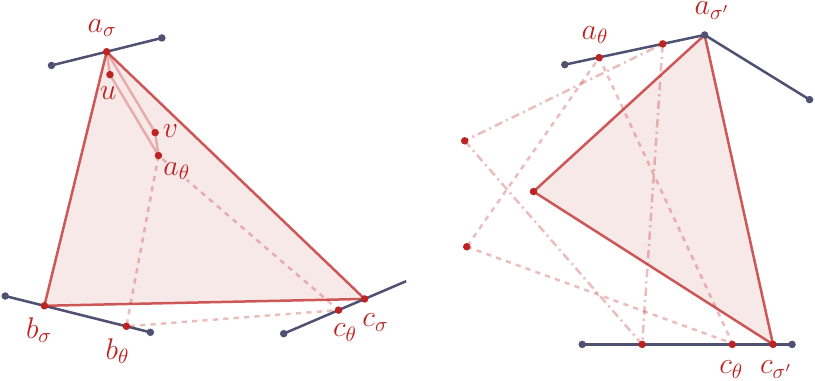}}
  \caption{Instances of Corollary~\ref{cor:placement_ac} and Lemma~\ref{lem:placement_ar}.}
  \label{fig:no_placement}
 \end{figure}

\begin{lemma}
\label{lem:placement_ar}
 For two consecutive $3$-pivots $\sigma$ and $\sigma'$ where
 the type of step $\PP[\sigma, \sigma']$ is $a_{r}$,
 there exists no orientation $\theta \in [0, 2\pi]$ such that
 both corners $b_{\theta}$ and $c_{\theta}$ of $\PP(\theta)$
 lie on $\partial Q(b_{\sigma}, c_{\sigma})$
 or $\partial Q(b_{\sigma'}, c_{\sigma'})$.
\end{lemma}
\begin{proof}
  Suppose for contradiction that there exists
  an orientation $\theta \in [0, 2\pi]$ where both corners $\btheta$ and~$\ctheta$
  lie on $\partial Q(b_{\sigma}, c_{\sigma})$.
  If the four corners $b_{\sigma}$, $\ctheta$, $\btheta$, and $c_{\sigma}$
  appear in a counter-clockwise order along $\partial Q$, then $\Ptheta$ can be enlarged,
  which contradicts the definition of $\Ptheta$.
  Thus, the relative order of the inner two corners~$\btheta$, $\ctheta$ 
  must be reversed, meaning that
  the four corners $b_{\sigma}$, $\btheta$, $\ctheta$, and $c_{\sigma}$
  appear in a counterclockwise order along $\partial Q$.
  
  We now consider the set of all possible placements of corner $a'$
  such that there exists a triangle $\triangle a'b'c'$ where
  $(i)$ $\triangle a'b'c' \sim \triangle abc$ and
  $(ii)$ two vertices $b', c'$ lie on the two line segments
  $\overline{b_{\sigma}\btheta}, \overline{\ctheta c_{\sigma}}$, respectively.
  Then, this set will be the parallelogram $\square a_{\sigma}u\atheta v$
  which satisfies $\triangle a_{\sigma}b_{\sigma}u
  \sim \triangle c_{\sigma}b_{\sigma}\ctheta$
  and $\triangle a_{\sigma}c_{\sigma}v \sim
  \triangle b_{\sigma}c_{\sigma}\btheta$.
  By the similarity of triangles, we obtain $\angle b_{\sigma}a_{\sigma}u
  = \angle b_{\sigma}c_{\sigma}\ctheta$
  and $\angle c_{\sigma}a_{\sigma}v
  = \angle c_{\sigma}b_{\sigma}\btheta$
  (see Figure~\ref{fig:no_placement}).
  
  From $\TT[\sigma, \sigma'] = a_{r}$,
  we obtain $(1)$ $\angle a + \phi_{bc}(\sigma) > \pi$
  when $c_{\sigma} \neq q_{c(\sigma)}$,
  and $(2)$ $\angle a + \phi(b(\sigma), c(\sigma)-1) > \pi$ 
  when $c_{\sigma} = q_{c(\sigma)}$.
  Since four corners $b_{\sigma}, \btheta, \ctheta, c_{\sigma}$
  appear in a counterclockwise order along $\partial Q$, by the convexity of $Q$,
  $(1)$ $\phi(\overrightarrow{\ctheta c_{\sigma}}, \overrightarrow{\btheta b_{\sigma}})
  \geq \phi_{bc}(\sigma)$ when $c_{\sigma} \neq q_{c(\sigma)}$ and
  $(2)$ $\phi(\overrightarrow{\ctheta c_{\sigma}}, \overrightarrow{\btheta b_{\sigma}})
  \geq \phi(b(\sigma), c(\sigma)-1)$ when~$c_{\sigma} = q_{c(\sigma)}$.
  In either case, we have
  $\phi(\overrightarrow{\ctheta c_{\sigma}}, \overrightarrow{\btheta b_{\sigma}})
  > \pi - \angle a$.
  
  The inequality with respect to $\angle b_{\sigma}a_{\sigma}c_{\sigma}$
  can be computed as follows:
  \begin{equation*}
   \angle b_{\sigma}a_{\sigma}c_{\sigma} = \angle a >
   \pi -\phi(\overrightarrow{\ctheta c_{\sigma}}, \overrightarrow{\btheta b_{\sigma}})
   = \pi - (\pi - \angle b_{\sigma}c_{\sigma}\ctheta
   - \angle c_{\sigma}b_{\sigma}\btheta)
   = \angle b_{\sigma}a_{\sigma}u + \angle c_{\sigma}a_{\sigma}v
  \end{equation*}
  With the equation $\overrightarrow{a_{\sigma}\atheta} = 
  \overrightarrow{a_{\sigma}u} + \overrightarrow{a_{\sigma}v}$,
  this inequality implies that the corner $\atheta$
  must lie in the interior of $\PP_{\sigma}$.  
  Since both corners $b_{\sigma}$ and $c_{\sigma}$ lie on $\partial Q(\ctheta, \btheta)$,
  by Corollary~\ref{cor:placement_ac},
  the placement of $\Ptheta$ cannot lie on the type $b_{c}$ or $c_{b}$ step,
  in other words, the corner $\atheta$ must lie on $\partial Q$.
  Therefore, the existence of orientation $\theta$ contradicts the definition of $\Ptheta$.
  The same arguement applies to the $3$-pivot $\sigma'$ by replacing $\sigma$ with $\sigma'$.
\end{proof}

\section{Steps and strides}
\label{sec:stride}

Our goal is to partition the global perfect rotation~$\PP[0, 2\pi]$
into a linear number of perfect rotations, each called \emph{stride}.
To introduce this specific perfect rotation,
we define the \emph{next head pivot} as follows:
\begin{definition}
  \label{def:next_head_pivot}
  For a given pivot $\sigma \in [0, 2\pi]$, the \emph{next head pivot} of $\sigma$
  is the smallest pivot~$\sigma''$ strictly larger than $\sigma$ that satisfies
  one of the followings:
  \begin{itemize}
   \item $\sigma''$ is a $2$-pivot,
   \item $\EE(\sigma'')$ differs from $\EE(\sigma)$ in at least two elements.
  \end{itemize}
\end{definition}
\noindent Beyond the definition of step, we define an extended step
using the next head pivot~$\sigma''$ of~$\sigma$.
\begin{definition}
  \label{def:stride}
  A \emph{stride} is a perfect rotation~$\PP[\sigma, \sigma'']$
  where~$\sigma$ is a pivot and~$\sigma''$ is
  the next head pivot of~$\sigma$.
\end{definition}
In other words, a stride is a sequence of steps of~$\Lambda$
such that all internal pivots of the stride are~$3$-pivots,
and the triples~$\Etheta$ of the steps vary only in one position.
While taking steps within a stride,
one element of~$\Etheta$ will change first
and only this position can be updated during the stride
up to the next head pivot~$\sigma''$.

The triple~$\Etheta$ can change on every step of a stride,
and the type of step can change as well.
To denote the types of steps within a single stride,
we define the type of the perfect rotation consist of multiple steps
as the sequence of types that appear on its steps:
\begin{definition}
  \label{def:type_multiple}
  For $k > 2$ and a perfect rotation~$\PP[\sigma_{1}, \sigma_{k}]$
  with two pivots~$\sigma_{1}$ and~$\sigma_{k}$, if there exist 
  pivots~$\sigma_{2}, \ldots, \sigma_{k-1} \in [\sigma_{1}, \sigma_{k}]$
  such that~$\sigma_{2} < \ldots < \sigma_{k-1}$ and
  the type of step changes from~$t_{\ell-1}$ to
  a different type~$t_{\ell} \neq t_{\ell-1}$
  at each pivot~$\sigma_{\ell}$ for~$\ell \in [2, k-1]$,
  then the \emph{type of the perfect rotation} is the string of
  types~$\TT[\sigma_{1}, \sigma_{k}] = t_{1}t_{2} \ldots t_{k-1}$.
\end{definition}
Since only one element of~$\Etheta$ can change during a stride,
the number of all possible types of stride is relatively small.
Thus, we will enumerate all possible types of stride,
which will lead us to a key property of the stride.
Before discussing the types of strides,
we first prove the following lemma about
a perfect rotation with specific type patterns.
\begin{lemma}
 \label{lem:type_abba}
 For a perfect rotation~$\PP[\sigma, \tau]$
 with two~$3$-pivots~$\sigma$ and $\tau$,
 if the type of the perfect rotation~$\TT[\sigma, \tau]$ is
 a string of combination of type~$a_{b}$ and type~$b_{a}$,
 then~$c(\sigma) \leq c(\tau)$.
 In particular, if the perfect rotation $\PP[\sigma, \tau]$
 satisfies one of the followings:
 \begin{itemize}
  \item $\PP[\sigma, \tau]$ is a step,
  \item the first step of $\PP[\sigma, \tau]$ is of type~$a_{b}$, or
  the last step of $\PP[\sigma, \tau]$ is of type~$b_{a}$, 
  \item the type of step changes from~$a_{b}$ to~$b_{r}$ at the pivot $\tau$,
 \end{itemize}
 then, we obtain $c(\sigma) < c(\tau)$.
\end{lemma}
\begin{proof}
 Since the pivot $\sigma$ is $3$-pivot,
 the type of step changes either $(1)$ from $a_{r}$ or $a_{c}$ to $a_{b}$, 
 or $(2)$ from $f_{w}$ or $b_{c}$ to $b_{a}$ at the pivot $\sigma$.
 Symmetrically, since $\tau$ is also a $3$-pivot,
 the type of step changes either $(1)$ from $a_{b}$ to $f_{w}$, $a_{c}$, or $b_{r}$,
 or $(2)$ from $b_{a}$ to $b_{r}$ or $b_{c}$ at the pivot $\tau$.
 All internal pivots of the perfect rotation $\PP[\sigma, \tau]$ are $2$-pivots,
 where the type of step changes either from $a_{b}$ to $b_{a}$,
 or from $b_{a}$ to $a_{b}$.

 If $\PP[\sigma, \tau]$ is a step,
 we have $\TT[\sigma, \tau] = a_{b}$ or~$b_{a}$.
 Since two pivots~$\sigma$ and~$\tau$ are~$3$-pivots,
 corner~$\ctheta$ departs from~$\partial Q$ at edge~$e_{c(\sigma)}$
 and returns to~$\partial Q$ again at edge~$e_{c(\tau)}$
 during its counter-clockwise motion.
 Then, the angle~$\angle c_{\tau}a_{\sigma}c_{\sigma} > 0$
 if~$\TT[\sigma, \tau] = a_{b}$, and
 the angle~$\angle c_{\tau}b_{\sigma}c_{\sigma} > 0$
 if~$\TT[\sigma, \tau] = b_{a}$.
 In both cases, we obtain~$c(\sigma) < c(\tau)$.
 
 We now assume that $\PP[\sigma, \tau]$ consists of multiple steps
 and let $\PP[\sigma, \sigma']$ and $\PP[\tau^{0}, \tau]$
 be the first and last step of $\PP[\sigma, \tau]$, respectively.
 
 When $\TT[\sigma, \sigma'] = a_{b}$,
 we have $\angle a + \phi_{bc}(\sigma) > \pi$.
 If $\TT[\tau^{0}, \tau] = a_{b}$, corner $\ctheta$ returns to $\partial Q$
 which implies that $\angle a + \phi_{bc}(\tau) < \pi$.
 Otherwise, $\TT[\tau^{0}, \tau] = b_{a}$
 which also implies that $\angle a + \phi_{bc}(\tau) < \pi$
 since $\angle b + \phi_{ca}(\tau) > \pi$.
 However, if $c(\sigma) \geq c(\tau)$, we obtain
 $\phi_{bc}(\sigma) < \phi_{bc}(\tau)$ since
 $b(\sigma) < b(\tau)$ and $Q$ is convex.
 This contradicts $\angle a + \phi_{bc}(\tau) < \pi
 < \angle a + \phi_{bc}(\sigma)$;
 thus, $c(\sigma) < c(\tau)$.
 
 When $\TT[\tau^{0}, \tau] = b_{a}$ or
 the type of step changes from $a_{b}$ to $b_{r}$ at the pivot $\tau$,
 we have $\angle b + \phi_{ca}(\tau) > \pi$.
 If $\TT[\sigma, \sigma'] = b_{a}$, corner $\ctheta$ departs from $\partial Q$,
 which implies that $\angle b + \phi_{ca}(\sigma) < \pi$.
 Otherrwise, the type of step changes from $a_{r}$ or $a_{c}$ to $a_{b}$,
 which also implies that $\angle b + \phi_{ca}(\sigma) < \pi$.
 However, if $c(\sigma) \geq c(\tau)$, we obtain
 $\phi_{ca}(\tau) \leq \phi_{ca}(\sigma)$ since
 $a(\sigma) \leq a(\tau)$ and $Q$ is convex.
 This contradicts $\angle b + \phi_{ca}(\sigma) < \pi
 < \angle b + \phi_{ca}(\tau)$;
 thus, $c(\sigma) < c(\tau)$.
 
 The remaining types of the perfect rotation~$\PP[\sigma, \tau]$
 are strictly alternating sequences of~$b_{a}$ and~$a_{b}$ of even length,
 i.e.~$\TT[\sigma, \tau] \in \{b_{a}a_{b}, b_{a}a_{b}b_{a}a_{b}, \ldots\}$.
 
 Let~$\alpha'$ be the smallest~$2$-pivot strictly larger than~$\sigma$
 such that the type of step changes from $b_{a}$ to $a_{b}$, and
 let~$\EE(\sigma) = (i, j, k)$ and~$\EE(\alpha') = (i', j, k)$.
 Then the corner~$\ctheta$ will move along
 the restricted boundary~$\partial Q[a_{\sigma}, a_{\alpha'}]$
 rotated clockwise by~$\angle b$ and
 scaled by~$|bc|/|ba|$ around~$b_{\sigma}$.
 Let~$c_{\sigma}, c_{i+1}, \ldots, c_{i'}$ be the vertices of
 this transformed boundary in the counter-clockwise order,
 where each~$c_{\ell}$ corresponds to~$q_{\ell}$ for~$\ell \in [i+1, i']$.
 By the definition, the angle~$\phi(\overrightarrow{c_{\sigma}c_{i+1}},
 \overrightarrow{q_{i}q_{i+1}}) = \angle b$
 and if~$i' \geq i+2$, the angle~$\phi(\overrightarrow{c_{\ell}c_{\ell+1}},
 \overrightarrow{q_{\ell}q_{\ell+1}}) = \angle b$ for $\ell \in [i+1, i'-1]$.
 
 Let~$\beta'$ be the smallest~$2$-pivot strictly larger than~$\alpha'$
 such that the type of step changes from~$a_{b}$ to~$b_{a}$
 or the corner~$c_{\beta'}$ returns to~$\partial Q$
 and let~$b(\beta') = j'$ (possibly $j' = j$).
 Then the corner~$\ctheta$ will move along
 the restricted boundary~$\partial Q[b_{\alpha'}, b_{\beta'}]$
 rotated counter-clockwise by~$\angle a$ and
 scaled by~$|ac|/|ab|$ around~$a_{\alpha'}$.
 Let~$c_{j}, \ldots, c_{j'}, c_{\beta'}$ be the vertices of
 this transformed boundary in the counter-clockwise order,
 where $c_{j} = c_{i'}$ and 
 each~$c_{\ell}$ corresponds to~$q_{\ell}$ for~$\ell \in [j, j']$
 (possibly~$c_{j'} = c_{\beta'}$).
 By the definition, if~$j' \geq j+1$,
 the angle~$\phi(\overrightarrow{q_{\ell}q_{\ell+1}},
 \overrightarrow{c_{\ell}c_{\ell+1}}) = \angle a$
 for~$\ell \in [j, j'-1]$
 and if $c_{j'} \neq c_{\beta'}$, 
 the angle~$\phi(\overrightarrow{q_{j'}q_{j'+1}},
 \overrightarrow{c_{j'}c_{\beta'}}) = \angle a$
 (see Figure~\ref{fig:two_steps_temp}).
 
 \begin{figure}
  \centerline{\includegraphics[width=0.5\textwidth]{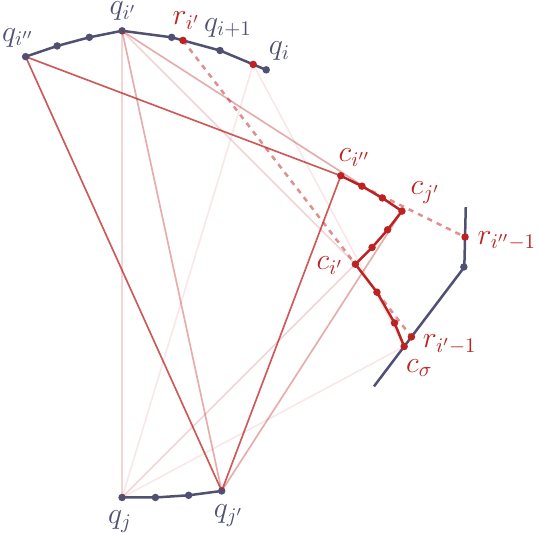}}
  \caption{The type $a_{b}$ and $b_{a}$ steps from pivot $\sigma$.}
  \label{fig:two_steps_temp}
 \end{figure}
 
 With these assumptions, for~$\ell \in [j, j'-1]$, we can compute
 the angle~$\phi(\overrightarrow{c_{i'}c_{i'-1}},
 \overrightarrow{c_{\ell}c_{\ell+1}})$ as follows:
 \begin{align*}
  \phi(\overrightarrow{c_{i'}c_{i'-1}}, \overrightarrow{c_{\ell}c_{\ell+1}})
  = \phi(\overrightarrow{c_{i'-1}c_{i'}}, \overrightarrow{c_{\ell+1}c_{\ell}})
  &= \phi(\overrightarrow{q_{i'-1}q_{i'}}, \overrightarrow{q_{\ell+1}q_{\ell}})
  + \angle a + \angle b \\
  &< \phi(\overrightarrow{q_{i'-1}q_{i'}}, \overrightarrow{q_{\ell+1}q_{\ell}})
  + \angle q_{i'-1}q_{i'}q_{\ell} + \angle q_{i'}q_{\ell}q_{\ell+1} = \pi
 \end{align*}
 This implies that the vertices~$c_{j}, \ldots, c_{j'}, c_{\beta'}$
 lie on the right open half-plane with respect to~$\overrightarrow{c_{i'-1}c_{i'}}$.
 Suppose that the line~$c_{i'-1}c_{i'}$ intersects~$\partial Q$ at~$r_{i'-1}$ and~$r_{i'}$,
 where~$r_{i'}$ is closer to~$c_{i'}$ than to~$c_{i'-1}$.
 Clearly,~$r_{i'-1}$ cannot lie on~$\partial Q(b_{\sigma}, c_{\sigma})$.
 Thus, if the corner~$c_{\beta'}$ returns to~$\partial Q$,
 the corner~$c_{\beta'}$ must lie on~$\partial Q(r_{i'-1}, r_{i'})$
 and therefore,~$c(\sigma) \leq c(\beta') = c(\tau)$.
 
 Otherwise, the corner~$c_{\beta'} = c_{j'}$ lies in the interior of~$Q$.
 If we define~$\alpha'', \beta'', i'', j''$ in the same way,
 the point~$r_{i''-1}$ cannot lie on~$\partial Q(b_{\sigma}, r_{i'-1})$
 since~$\phi(\overrightarrow{c_{i'-1}c_{i'}},
 \overrightarrow{c_{i''-1}c_{i''}}) > 0$.
 So, if the corner~$c_{\beta''}$ returns to~$\partial Q$,
 we obtain~$c(\sigma) < c(\beta'') = c(\tau)$, again.
 Therefore, by repeating this process until the corner~$\ctheta$ returns to~$\partial Q$,
 we always have~$c(\sigma) \leq c(\tau)$.
\end{proof}

Lemma~\ref{lem:type_abba} remains valid when~$a, b, c$ are cyclically rotated.
With Definition~\ref{def:type_multiple} and Lemma~\ref{lem:type_abba},
we have the following important lemma.

\begin{lemma}
  \label{lem:prop_stride}
  A stride contains at most five different types of steps and after passing through
  a stride, at least one element of~$\Etheta$ increases.
\end{lemma}
\begin{proof}
 For a given pivot~$\sigma$, let~$\sigma'$ and $\sigma''$ be
 the next pivot and next head pivot of~$\sigma$, respectively.
 If~$\sigma'$ is $2$-pivot,
 by the definition, we have $\sigma' = \sigma''$.
 Then the chain~$\PP[\sigma, \sigma'']$ is a step of~$\Lambda$
 with one of six types~$a_{b}, a_{c}, b_{c}, b_{a}, c_{a}, c_{b}$.
 Thus, after passing through this step,
 one element of~$\Etheta$ increases by one.

 Suppose that $\sigma'$ is $3$-pivot.
 Depending on the type of the first step~$\TT[\sigma, \sigma']$,
 we enumerate all possible types of strides~$\TT[\sigma, \sigma'']$
 and compare $\EE(\sigma'')$ with~$\EE(\sigma)$.

\begin{enumerate}
 \item When~$\TT[\sigma, \sigma'] = f_{w}$,
 the first two pivots~$\sigma$ and $\sigma'$ are $3$-pivots.
 Without loss of generality, we assume that
 $a(\theta)$ increases by one at the pivot~$\sigma'$,
 which implies that only~$a(\theta)$ can vary
 throughout the perfect rotation~$\PP[\sigma, \sigma'')$.
 Then we can find all possible~$\TT[\sigma, \sigma'']$
 by taking red-colored arrows from state~$f_{w}$
 and a blue-colored arrow from~$c_{r}$ to~$c_{a}$ from Figure~\ref{fig:state_diagram}.
 
 \begin{enumerate}
  \item $\TT[\sigma, \sigma''] = f_{w}$:
  the last step of stride is of type~$f_{w}$; the stride terminates
  when either the corner~$b_{\sigma''}$ coincides with vertex~$q_{b(\sigma)+1}$
  or the corner~$c_{\sigma''}$ coincides with vertex~$q_{c(\sigma)+1}$.
  Thus, $(1)$~$a(\sigma'') > a(\sigma)$ and 
  $(2)$ either $b(\sigma'') = b(\sigma) + 1$ or~$c(\sigma'') = c(\sigma) + 1$.
  
  \medskip
  \item $\TT[\sigma, \sigma''] = f_{w}a_{c}$:
  there exists a~$3$-pivot~$\alpha^{(1)} \in [\sigma', \sigma'')$
  such that the type of step changes from~$f_{w}$ to~$a_{c}$.
  Then, the corner $a_{\alpha^{(1)}}$ coincides with
  the vertex $q_{a(\alpha^{(1)})}$.
  
  The last step of stride is of type~$a_{c}$;
  either the corner~$b_{\sigma''}$ returns to~$\partial Q$ or
  the corner~$c_{\sigma''}$ coincides with
  vertex~$q_{c(\sigma)+1}$ at the $2$-pivot~$\sigma''$.
  Since all corners of~$\Ptheta$ move forward
  during the perfect rotation~$\PP[\sigma, \alpha^{(1)}]$,
  it follows that $(1)$~$a(\sigma'') > a(\sigma)$ and
  $(2)$ either~$b(\sigma'') > b(\sigma)$ by Lemma~\ref{lem:type_abba},
  or~$c(\sigma'') = c(\sigma) + 1$.
  
  \medskip
  \item When there exists a $3$-pivot $\alpha^{(2)} \in [\sigma', \sigma'')$
  such that the type of step changes from~$f_{w}$ to~$c_{r}$.
  Then, the corner $a_{\alpha^{(2)}}$ coincides with
  the vertex $q_{a(\alpha^{(2)})}$.
  
  \begin{enumerate}
   \item $\TT[\sigma, \sigma''] = f_{w}c_{r}$:
   the last step of stride is of type~$c_{r}$;
   either $(1)$ the corner~$b_{\sigma''}$ coincides with the vertex~$q_{b(\sigma)+1}$
   or $(2)$ the corner~$a_{\sigma''}$ coincides with a vertex of~$Q$
   and~$c(\sigma'') = c(\sigma) - 1$.
   In either case, we obtain $a(\sigma'') > a(\sigma)$ since
   corner~$\atheta$ always moves forward throughout the stride.
   
   \smallskip
   \item $\TT[\sigma, \sigma''] = f_{w}c_{r}a_{c}$:
   there exists another $3$-pivot~$\alpha^{(3)} \in (\alpha^{(2)}, \sigma'')$ such that
   the type of step changes from~$c_{r}$ to~$a_{c}$.
   Then, the corner $a_{\alpha^{(3)}}$ coincides with the vertex $q_{a(\alpha^{(3)})}$.
   
   The last step of stride is of type~$a_{c}$;
   either the corner~$b_{\sigma''}$ returns to~$\partial Q$ or
   the corner~$c_{\sigma''}$ coincides with
   vertex~$q_{c(\sigma)+1}$ at the $2$-pivot~$\sigma''$.
   Thus, it follows that $(i)$~$a(\sigma'') > a(\sigma)$ 
   since the corner~$\atheta$ always moves forward
   during the perfect rotation~$\PP[\sigma, \alpha']$
   and~$(ii)$ either $b(\sigma'') > b(\sigma)$ by Lemma~\ref{lem:type_abba},
   or $c(\sigma'') = c(\sigma) + 1$.

   \smallskip
   \item $\TT[\sigma, \sigma''] = f_{w}c_{r}c_{a}$:
   there exists another $3$-pivot $\gamma^{(1)} \in (\alpha^{(2)}, \sigma'')$ such that
   the type of step changes from $c_{r}$ to $c_{a}$.
   This happens when corner $\ctheta$ moves backward along the edge $e_{c(\sigma)}$,
   and corner $c_{\gamma^{(1)}}$ coincides with the vertex $q_{c(\sigma)}$.
   
   The last step of stride is of type $c_{a}$; either
   the corner $a_{\sigma''}$ coincides with
   the vertex $q_{a(\gamma^{(1)})+1}$ at the $2$-pivot $\sigma''$
   or the corner $b_{\sigma''}$ returns to $\partial Q$.
   In either case, we have $a(\sigma'') > a(\sigma)$ since
   the corner $\atheta$ always moves forward throughout the stride.
  \end{enumerate}
 \end{enumerate}
 
 \medskip
 Let $\TT_{a} = \{\varepsilon, a_{c}, c_{r}, c_{r}a_{c}, c_{r}c_{a}\}$ be a set of strings 
 and define two sets $\TT_{b}$ and $\TT_{c}$, cyclically.
 Then, all possible types $\TT[\sigma, \sigma'']$
 when $\TT[\sigma, \sigma'] = f_{w}$ and $a(\theta)$ is updated first
 at the next pivot $\sigma'$
 can be expressed as the concatenation $f_{w}s$ for each $s \in \TT_{a}$.
 
 In general, let $\TT[f_{w}]$ be the set of all possible types $\TT[\sigma, \sigma'']$
 when $\TT[\sigma, \sigma'] = f_{w}$.
 Then, we can express the set~$\TT[f_{w}] = \{f_{w}\} \cup 
 \{f_{w}s \mid s \in \TT_{a} \cup \TT_{b} \cup \TT_{c}\}$.
 
 \bigskip
 \item When $\TT[\sigma, \sigma'] = a_{c}$,
 corner $\btheta$ returns to $\partial Q$ at the $3$-pivot $\sigma'$
 and $b(\sigma) < b(\sigma')$ by Lemma~\ref{lem:type_abba}.
 Since only $b(\theta)$ can change throughout the stride,
 we can find all $\TT[\sigma, \sigma'']$ by taking green-clored arrows
 from state $a_{c}$ and a red-colored arrow from $a_{r}$ to $a_{b}$ from Figure~\ref{fig:state_diagram}.

 \begin{enumerate}
  \item $\TT[\sigma, \sigma''] = a_{c}a_{b}$:
  Assume that the type of step changes from $a_{c}$ to $a_{b}$
  at $3$-pivot $\sigma'$.
  
  The last step of stride is of type $a_{b}$;
  either the corner $b_{\sigma''}$ coincides with
  the vertex $q_{b(\sigma')+1}$ at $2$-pivot $\sigma''$
  or the corner $c_{\sigma''}$ returns to $\partial Q$.
  In either case, we have $b(\sigma'') > b(\sigma)$.

  \medskip
  \item $\TT[\sigma, \sigma''] = a_{c}a_{r}$:
  assume that the type of step changes from $a_{c}$ to $a_{r}$
  at $3$-pivot $\sigma'$.
  
  The last step of stride is of type $a_{r}$;
  either $(1)$ the corner $b_{\sigma''}$ or $c_{\sigma''}$
  coincides with a vertex of $Q$
  and $a(\sigma'') = a(\sigma) - 1$ or
  $(2)$ the corner $a_{\sigma''}$ coincides with
  vertex $q_{a(\sigma)-1}$.
  In either case, we have $b(\sigma'') > b(\sigma)$ since corner $\btheta$
  moves forward during the perfect rotation $\PP[\sigma', \sigma'']$.
 \end{enumerate}
 
 \medskip
 Let $\TT[a_{c}]$ be the set of all possible types $\TT[\sigma, \sigma'']$
 when $\TT[\sigma, \sigma'] = a_{c}$.
 Then, we can express the set~$\TT[a_{c}] = \{a_{c}, a_{c}a_{b}, a_{c}a_{r}\}$.
 
 \bigskip
 \item When $\TT[\sigma, \sigma'] = a_{b}$,
 corner $c_{\sigma'}$ returns to $\partial Q$ at the $3$-pivot $\sigma'$
 and $c(\sigma) \leq c(\sigma')$ by Lemma~\ref{lem:type_abba}.

 \begin{enumerate}
  \item $\TT[\sigma, \sigma''] = \{a_{b}s \mid s \in \TT[a_{c}]\}$:
  assume that the type of step changes from $a_{b}$ to $a_{c}$
  at $3$-pivot $\sigma'$.
  
  If $c(\sigma) < c(\sigma')$, either the corner $b_{\sigma''}$ returns to $\partial Q$
  or the corner $c_{\sigma''}$ coincides with the vertex $q_{c(\sigma')+1}$
  at the $2$-pivot $\sigma''$. Thus, $\TT[\sigma, \sigma''] = a_{b}a_{c}$
  and $c(\sigma'') > c(\sigma)$.  
  Otherwise, $c(\sigma) = c(\sigma')$,
  which reduces to the situation when $\TT[\sigma, \sigma'] = a_{c}$;
  thus, $\TT[\sigma, \sigma''] \in \{a_{b}a_{c}a_{b}, a_{b}a_{c}a_{r}\}$
  and $b(\sigma'') > b(\sigma)$.

  \medskip
  \item $\TT[\sigma, \sigma''] = \{a_{b}s \mid s \in \TT[f_{w}]\}$:
  assume that the type of step changes from $a_{b}$ to $f_{w}$ at $3$-pivot~$\sigma'$.
  
  If $c(\sigma) < c(\sigma')$, the case reduces to the situation
  when $\TT[\sigma, \sigma'] = f_{w}$ and $c(\theta)$ is updated first.
  Otherwise, $c(\sigma) = c(\sigma')$, the case reduces to the situation
  when $\TT[\sigma, \sigma'] = f_{w}$.
  By the observation of $\TT[f_{w}]$, at least one element of $\Etheta$ increases.
  
  \medskip
  \item Assume that the type of step changes from $a_{b}$ to $b_{r}$ at $3$-pivot $\sigma'$.
  Then, we obtain $c(\sigma) < c(\sigma')$ by Lemma~\ref{lem:type_abba}
  and only $c(\theta)$ can change throughout the stride.
  \begin{enumerate}
   \item $\TT[\sigma, \sigma''] = a_{b}b_{r}$: 
   the last step of stride is of type $b_{r}$;
   either $(1)$ the corner $a_{\sigma''}$ coincides with the vertex $q_{a(\sigma)+1}$ or
   $(2)$ the corner $c_{\sigma''}$ coincides with a vertex of $Q$ and
   $b(\sigma'') = b(\sigma) - 1$.
   In either case, we have $c(\sigma'') > c(\sigma)$ since
   corner $\ctheta$ moves forward during the perfect rotation $\PP[\sigma, \sigma'']$.
   
   \smallskip
   \item $\TT[\sigma, \sigma''] = a_{b}b_{r}c_{b}$:
   there exists another $3$-pivot $\gamma^{(1)} \in (\sigma', \sigma'')$ such that
   the type of step changes from $b_{r}$ to $c_{b}$.
   
   The last step of stride is of type $c_{b}$;
   either the corner $a_{\sigma''}$ returns to $\partial Q$ or
   the corner $b_{\sigma''}$ coincides with
   the vertex $q_{b(\sigma)+1}$ at the $2$-pivot $\sigma''$.
   Thus, it follows that $(i)$ $c(\sigma'') > c(\sigma)$ and
   $(ii)$ either $a(\sigma'') > a(\sigma)$ by Lemma~\ref{lem:type_abba}, or
   $b(\sigma'') = b(\sigma) + 1$.
   
   \smallskip
   \item $\TT[\sigma, \sigma''] = a_{b}b_{r}b_{c}$:
   there exists another $3$-pivot $\beta^{(1)} \in (\sigma', \sigma'')$ such that
   the type of step changes from $b_{r}$ to $b_{c}$.
   
   The last step of stride is of type $b_{c}$;
   either the corner $a_{\sigma''}$ returns to $\partial Q$ or
   the corner $c_{\sigma''}$ coincides with
   the vertex $q_{c(\beta^{(1)})+1}$ at the $2$-pivot $\sigma''$.
   In either case, we have $c(\sigma'') > c(\sigma)$ since
   corner $\ctheta$ moves forward during the perfect rotation $\PP[\sigma', \sigma'']$.
  \end{enumerate}
  
  \smallskip
  In short, we have
  $\TT[\sigma, \sigma''] \in \{a_{b}s \mid s \in \TT_{c}
  \setminus \{\varepsilon, c_{b}\}\}$
  where $\TT_{c} = \{\varepsilon, c_{b}, b_{r}, b_{r}c_{b}, b_{r}b_{c}\}$.
 \end{enumerate}
 \medskip
 Let $\TT[a_{b}]$ be the set of all possible types $\TT[\sigma, \sigma'']$
 when $\TT[\sigma, \sigma'] = a_{b}$.
 Then, we can express the 
 set~$\TT[a_{b}] = \{a_{b}s \mid s \in \TT[a_{c}] \cup \TT[f_{w}]
 \cup (\TT_{c} \setminus \{c_{b}\})\}$,
 where $\TT_{c} = \{\varepsilon, c_{b}, b_{r}, b_{r}c_{b}, b_{r}b_{c}\}$.
 
 \bigskip
 \item When $\TT[\sigma, \sigma'] = a_{r}$,
 two pivots $\sigma$ and $\sigma'$ are $3$-pivots.
 Depending on the first updated element of $\Etheta$,
 we can find all possible types of chain by following specific colored arrows
 and a red-colored arrow from $a_{r}$ to $a_{b}$ and
 green-colored arrow from $b_{r}$ to $b_{c}$ from Figure~\ref{fig:state_diagram}.

 \begin{enumerate}
  \item $\TT[\sigma, \sigma''] = a_{r}$:
  by the definition of the next head pivot $\sigma''$, the stride will be finished
  when two elements of $\Etheta$ change.
  Thus, either $b(\sigma'') > b(\sigma)$ or $c(\sigma'') > c(\sigma)$
  since two corners $\btheta, \ctheta$ move forward
  throughout the stride.
  
  \medskip
  \item $\TT[\sigma, \sigma''] = a_{r}b_{a}$ :
  there exists a $3$-pivot $\beta^{(1)} \in [\sigma', \sigma'')$
  such that the type of step changes from $a_{r}$ to $b_{a}$.
  Then, the corner $b_{\beta^{(1)}}$ coincides with vertex $q_{b(\beta^{(1)})}$
  and $b(\beta^{(1)}) > b(\sigma)$ since corner $\btheta$ moves forward
  during the perfect rotation $\PP[\sigma, \beta^{(1)}]$.
  
  The last step of stride is of type $b_{a}$;
  the corner $a_{\sigma''}$ coincides with
  the vertex $q_{a(\sigma)+1}$ at $2$-pivot~$\sigma''$
  or the corner $c_{\sigma''}$ returns to $\partial Q$.
  Thus, we have $(i)$ $b(\sigma'') > b(\sigma)$
  and $(ii)$ either $a(\sigma'') = a(\sigma) + 1$ or $c(\sigma'') > c(\sigma)$
  by Lemma~\ref{lem:type_abba}.
  
  \medskip
  \item Assume that there exists a $3$-pivot $\alpha^{(1)} \in [\sigma', \sigma'')$
  such that the type of step changes from $a_{r}$ to $a_{b}$.
  
  \begin{enumerate}
   \item $\TT[\sigma, \sigma''] = a_{r}a_{b}$:
   the last step of stride is of type $a_{b}$;
   if $a(\sigma'') = a(\sigma)$, the corner $b_{\sigma''}$
   coincides with the vertex $q_{b(\sigma)+1}$ at the $2$-pivot $\sigma''$.
   Otherwise, $a(\sigma'') < a(\sigma)$, the corner $c_{\sigma''}$
   also can return to $\partial Q$.
   Thus, we have either $b(\sigma'') > b(\sigma)$ or
   $c(\sigma'') > c(\sigma)$ by Lemma~\ref{lem:type_abba}.
   
   \smallskip
   \item $\TT[\sigma, \sigma''] \in \{a_{r}a_{b}s \mid
   s \in \TT_{c}\setminus \{\varepsilon, c_{b}\}\}$:
   there exists another $3$-pivot $\gamma^{(1)} \in (\alpha^{(1)}, \sigma'']$
   such that the type of step changes from $a_{b}$ to $b_{r}$.
   Then, the case reduces to the situation $3$-$(c)$ when $\TT[\sigma, \sigma'] = a_{b}$.
   By the observation of $3$-$(c)$, at least one element of $\Etheta$ increases.
   
   \smallskip
   \item $\TT[\sigma, \sigma''] = a_{r}a_{b}a_{c}$:
   there exists another $3$-pivot $\gamma^{(2)} \in (\alpha^{(1)}, \sigma'']$
   such that the type of step changes from $a_{b}$ to $a_{c}$.
   Then, the corner $c_{\gamma^{(1)}}$ returns to $\partial Q$
   and $c(\gamma^{(2)}) > c(\alpha^{(1)})$
   by Lemma~\ref{lem:type_abba}.
   
   The last step of stride is of type $a_{c}$;
   the corner $b_{\sigma''}$ returns to $\partial Q$ or
   the corner $c_{\sigma''}$ coincides with
   the vertex $q_{c(\gamma^{(2)})+1}$ at $2$-pivot $\sigma''$.
   In either case, we have $c(\sigma'') > c(\sigma)$
   since corner $\ctheta$ moves forward
   during the perfect rotation $\PP[\sigma, \alpha^{(1)}]$.
   
   \smallskip
   \item $\TT[\sigma, \sigma''] \in \{a_{r}a_{b}f_{w}s \mid s \in \TT_{c}\}$:
   there exists another $3$-pivot $\gamma^{(3)} \in (\alpha^{(1)}, \sigma'']$
   such that the type of step changes from $a_{b}$ to $f_{w}$.
   Then, the corner $c_{\gamma^{(3)}}$ returns to $\partial Q$
   and $c(\gamma^{(3)}) > c(\alpha^{(1)})$
   by Lemma~\ref{lem:type_abba}.
   This means that the case reduces to the situation when $\TT[\sigma, \sigma'] = f_{w}$
   and $c(\theta)$ is updated first.
   By the observation of $\TT[f_{w}]$, at least one element of $\Etheta$ increases.
  \end{enumerate}

  \medskip
  \item $\TT[\sigma, \sigma''] \in \{a_{r}f_{w}s \mid s \in \TT_{c}\}$:
  assume that there exists a $3$-pivot $\gamma^{(4)} \in [\sigma', \sigma'')$
  such that the type of step changes from $a_{r}$ to $f_{w}$.
  Then, $c(\gamma^{(4)}) > c(\sigma)$ which reduces to the situation
  when $\TT[\sigma, \sigma'] = f_{w}$ and $c(\theta)$ is updated first.
  By the observation of $\TT[f_{w}]$, at least one element of $\Etheta$ increases.
  
  \medskip
  \item $\TT[\sigma, \sigma''] \in \{a_{r}s \mid
  s \in \TT_{c} \setminus \{\varepsilon, c_{b}\}\}$:
  assume that there exists a $3$-pivot $\gamma^{(5)} \in [\sigma', \sigma'')$
  such that the type of step changes from $a_{r}$ to $b_{r}$.
  Then, $c(\gamma^{(5)}) > c(\sigma)$ which reduces to the situation $3$-$(c)$
  when $\TT[\sigma, \sigma'] = a_{b}$.
  By the observation of $3$-$(c)$, at least one element of $\Etheta$ increases.
 \end{enumerate}
 
 \medskip
 Let $\TT[a_{r}]$ be the set of all possible types $\TT[\sigma, \sigma'']$
 when $\TT[\sigma, \sigma'] = a_{r}$. Then, we can express the set
 $\TT[a_{r}] = \{a_{r}b_{a}, a_{r}a_{b}a_{c}\}
 \cup \{a_{r}s \mid s \in \TT_{c} \setminus \{c_{b}\}\} 
 \cup \{a_{r}a_{b}s \mid s \in \TT_{c} \setminus \{c_{b}\}\}
 \cup \{a_{r}f_{w}s \mid s \in \TT_{c}\}
 \cup \{a_{r}a_{b}f_{w}s \mid s \in \TT_{c}\}$,
 where $\TT_{c} = \{\varepsilon, c_{b}, b_{r}, b_{r}c_{b}, b_{r}b_{c}\}$.
\end{enumerate}

 The remaining $\TT[b_{c}]$, $\TT[b_{a}]$, $\TT[b_{r}]$,
 $\TT[c_{a}]$, $\TT[c_{b}]$, $\TT[c_{r}]$ can be obatined
 by permuting $a, b, c$, cyclically.
 Since the length of string $\TT[\sigma, \sigma'']$ is at most five,
 a stride contains at most five diffrent types of steps
 and we have shown that at least one element of $\Etheta$ increases
 after passing the stride.
\end{proof}

Based on the fact that a stride has at most five types of steps,
we can compute a stride using only binary search as follows:

\begin{lemma}
  \label{lem:take_stride}
  Given a pivot~$\sigma$, the stride starting at~$\sigma$ can be
  computed in~$O(\log n)$ time.
\end{lemma}
\begin{proof} 
  Let~$\sigma'$ and~$\sigma''$ be the next pivot and the next head pivot
  of pivot~$\sigma$, respectively, and let $\TT[\sigma, \sigma'] = t_{0}$
  and $\EE(\sigma) = (i, j, k)$.
  Then, if $\TT[\sigma, \sigma''] \neq t_{0}$,
  we define $\sigma^{(1)} < \sigma''$ as the smallest pivot
  such that the type of step changes from $t_{0}$ to other type $t_{1}$
  at pivot $\sigma^{(1)}$.  We claim that we can compute
  the perfect rotation $\PP[\sigma, \sigma^{(1)}]$ and
  decide the type $t_{1}$ in $O(\log n)$ time if it exists.
 
  We compute the type~$t_{0}$ by computing~$\phi_{ab}(\sigma),
  \phi_{bc}(\sigma), \phi_{ca}(\sigma)$.
  
  \begin{itemize}
   \item When $t_{0} = a_{c}$
   (when $t_{0} = b_{a}, c_{b}, a_{b}, b_{c}, c_{a}$ are symmetric): 
   corner $\atheta$ is fixed at vertex $q_{i}$ and
   corner $\ctheta$ moves along edge $e_{k}$ in counter-clockwise direction.
   Thus, we compute the edge $e_{k}$ rotated clockwise direction by $\angle a$
   and scaled by $|ab|/|ac|$ from $\atheta$ and then,
   find the intersection between this edge and $\partial Q$ using binary search.
   
   If there exists an intersection point, this point will be a vertex $b_{\sigma^{(1)}}$
   and a vertex $a_{\sigma^{(1)}}$ will be $q_{i}$ so that
   we can compute the placement of $\PP_{\sigma^{(1)}}$. 
   By computing three angles $\phi_{ab}(\sigma^{(1)}),
   \phi_{bc}(\sigma^{(1)}), \phi_{ca}(\sigma^{(1)})$,
   we can decide the type $t_{1}$.
  
   Otherwise, there is no intersection point which means that
   corner $\ctheta$ coincides with vertex $q_{k+1}$ at $2$-pivot $\sigma''$,
   in other words, $\TT[\sigma, \sigma''] = a_{c}$.
 
   \item When $t_{0} = f_{w}$: we first take a step from the $3$-pivot $\sigma$
   and find which element of $\Etheta$ is updated first at the $3$-pivot $\sigma'$.
   Without loss of generality, we assume that $a(\theta)$ is updated first.
   Then, the type of step can change from $f_{w}$ to $c_{r}$ or $a_{c}$,
   we have three following cases:
   
   \begin{itemize}
    \item If pivot $\sigma^{(1)}$ does not exist,
    that is $\TT[\sigma, \sigma''] = f_{w}$, all three corners of $\Ptheta$
    moves forward during the stride $\PP[\sigma, \sigma'']$ and
    either the vertex $b_{\sigma''}$ or $c_{\sigma''}$ coincides with
    the vertex $q_{j+1}$ or $q_{k+1}$.
    
    To find the next head pivot $\sigma''$, we construct
    the parallelogram~$\square_{bc}(e_{j}, e_{k})$ which is
    the set of all possible placements of vertex~$a'$ such that
    there exists $\triangle a'b'c'$ where $\triangle a'b'c' \sim \triangle abc$
    and two vertices $b'$, $c'$ lie on two edges $e_{j}$, $e_{k}$, respectively
    (as we previously showed in Lemma~\ref{lem:placement_ar}).
    Then, for a point $a^{\ast}  \in \square_{bc}(e_{j}, e_{k}) \cap \partial Q$,
    there exist two vertices $b^{\ast}$, $c^{\ast}$ where
    $\triangle abc \sim \triangle  a^{\ast}b^{\ast}c^{\ast}$ and
    two vertices $b^{\ast}$, $c^{\ast}$ lie on two edges~$e_{j}$, $e_{k}$, respectively.
    In particular, if vertex $a^{\ast}$ lies on the boundary of~$\square_{bc}(e_{j}, e_{k})$,
    then either vertex $b^{\ast}$ or vertex $c^{\ast}$ coincides with a vertex of~$Q$
    (see Figure~\ref{fig:parallelogram}).
    
    \begin{figure}
     \centerline{\includegraphics[width=0.5\textwidth]{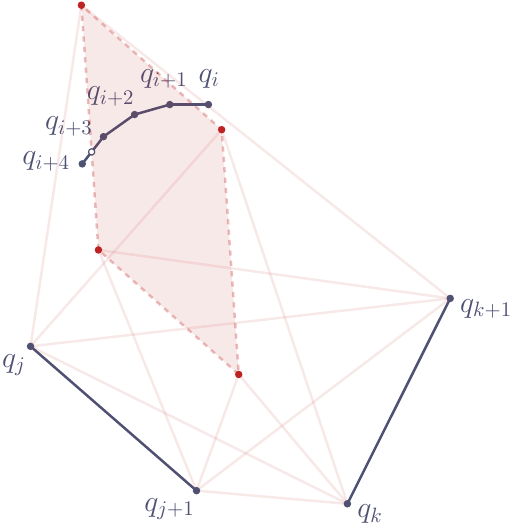}}
     \caption{The parallelogram $\square_{bc}(e_{j}, e_{k})$ in Lemma~\ref{lem:take_stride}.}
     \label{fig:parallelogram}
    \end{figure}
  
    Thus, we can find the vertex $a_{\sigma''}$ from the intersection
    between the boundary of $\square_{bc}(e_{j}, e_{k})$ and $\partial Q$
    using a binary search (if the intersection is not unique,
    we must pick the point that is the closest to vertex $a_{\sigma}$ along $\partial Q$).
    Two vertices $b_{\sigma''}$ or $c_{\sigma''}$ can also be found
    from the relative placement of $a_{\sigma''}$ 
    with respect to $\square_{bc}(e_{j}, e_{k})$;
    thus, the next head pivot~$\sigma''$ can be computed in $O(\log n)$ time.
    
    \item If the type of step changes from $t_{0} = f_{w}$
    to $t_{1} = c_{r}$, the corner $a_{\sigma^{(1)}}$ coincides with
    the vertex $q_{a(\sigma^{(1)})}$ and
    $\angle c + \phi_{ab}(\theta)$ becomes larger than $\pi$
    at the $3$-pivot $\sigma^{(1)}$.
    Since $\phi_{ab}(\theta)$ increases as corner $\atheta$ moves forward,
    we can find the $3$-pivot $\sigma^{(1)}$ in $O(\log n)$ time using a binary search.
    
    \item If the type of step changes from $t_{0} = f_{w}$
    to $t_{1} = a_{c}$, the corner $a_{\sigma^{(1)}}$ coincides with
    the vertex $q_{a(\sigma^{(1)})}$ and $\phi_{ca}(\theta)$ becomes negative
    at the $3$-pivot $\sigma^{(1)}$.
    Since $\phi_{ca}(\theta$) decreases as corner $\atheta$ moves forward,
    we can find the $3$-pivot $\sigma^{(1)}$ in $O(\log n)$ time using a binary search.
   \end{itemize}
   We compute a candidate pivot for each of the three cases,
   and determine the pivot that appears first.
   
   \item When $t_{0} = a_{r}$ (when $t_{0} = b_{r}, c_{r}$ are symmetric):
   the type of step can change from $a_{r}$ to $a_{b}$, $b_{a}$, or $f_{w}$,
   we have four following cases:
   \begin{itemize}
    \item If pivot $\sigma^{(1)}$ does not exist,
    that is $\TT[\sigma, \sigma''] = a_{r}$,
    two corners $\btheta$ and $\ctheta$ move forward while
    corner $\atheta$ moves backward.
    If $a(\theta)$ is updated first, construct
    the parallelogram $\square_{bc}(e_{j}, e_{k})$ and find the intersection
    between the boundary of $\square_{bc}(e_{j}, e_{k})$ and $\partial Q$
    using a binary search. 
    Then, we can find the next head pivot $\sigma''$ in $O(\log n)$ time.
    
    Without loss of generality, assume that $b(\theta)$ is updated first.
    Construct the parallelogram $\square_{ca}(e_{k}, e_{i})$ and
    find the intersection $b_{\sigma^{\ast}}$ between
    the boundary of $\square_{ca}(e_{k}, e_{i})$ and $\partial Q$
    using a binary search.
    If corner $c_{\sigma^{\ast}}$ coincides with vertex $q_{j+1}$,
    then $\sigma^{\ast} = \sigma''$.
    Otherwise, corner $a_{\sigma^{\ast}}$ coincides with vertex $q_{i}$,
    which means that $a(\sigma^{\ast}) = i$.
    Then we need to take one more step from the pivot $\sigma^{\ast}$
    to find the next head pivot $\sigma''$.
    
    \item If the type of step changes from $t_{0} = a_{r}$
    to $t_{1} = f_{w}$, the corner $c_{\sigma^{(1)}}$ coincides with
    the vertex $q_{c(\sigma^{(1)})}$ and
    $\angle a + \phi_{bc}(\theta)$ becomes equal or smaller than $\pi$
    at the $3$-pivot $\sigma^{(1)}$.
    Only when $c(\theta)$ is updated first,
    $\phi_{bc}(\theta)$ decreases as corner $\btheta$ moves forward.
    Thus, we can find the $3$-pivot $\sigma^{(1)}$ using a binary search.
 
    \item If the type of step changes from $t_{0} = a_{r}$
    to $t_{1} = a_{b}$, the corner $a_{\sigma^{(1)}}$ coincides with
    the vertex $q_{a(\sigma^{(1)})}$ and
    $\phi_{ab}(\theta)$ becomes negative at the $3$-pivot $\sigma^{(1)}$.
    When $a(\theta)$ or $b(\theta)$ is updated first,
    $\phi_{ab}(\theta)$ decreases as corner $\atheta$ moves backward
    or corner $\btheta$ moves forward.
    Thus, we can find the $3$-pivot $\sigma^{(1)}$ using a binary search.

    \item If the type of step changes from $t_{0} = a_{r}$
    to $t_{1} = b_{a}$, the corner $b_{\sigma^{(1)}}$ coincides with
    the vertex $q_{b(\sigma^{(1)})}$ and
    $\phi_{ab}(\theta)$ becomes negative at the $3$-pivot $\sigma^{(1)}$.
    When $a(\theta)$ or $b(\theta)$ is updated first,
    $\phi_{ab}(\theta)$ decreases as corner $\atheta$ moves backward
    or corner $\btheta$ moves forward.
    Thus, we can find the $3$-pivot $\sigma^{(1)}$ using a binary search.       
   \end{itemize}
    We compute a candidate pivot for each of the four cases,
   and determine the pivot that appears first.
  \end{itemize}
 
   If pivot $\sigma^{(1)}$ exists, then we find the pivot $\sigma^{(2)}$
   such that the type of step changes from $t_{1}$ to other type $t_{2}$ in
   a similar way, and so on.
   Since a stride has at most five types of steps,
   we need to use binary search at most five times.
\end{proof}

Lemma~\ref{lem:take_stride} also implies that
a stride consists of at most three chains and two steps with one of
types $a_{b}$, $a_{c}$, $b_{a}$, $b_{c}$, $c_{a}$, $c_{b}$.
With this lemma, we will present how to partition the global perfect rotation
into a linear number of strides in the following section.

\section{Partitioning the global perfect rotation}
\label{sec:final_partition}

Without loss of generality, suppose that the orientation $0$ is a $3$-pivot of $\Lambda$.
We then sequentially apply Lemma~\ref{lem:take_stride} from $\PP(0)$
until the global perfect rotation is entirely covered by strides.
Our goal is to prove that the total number of these strides is $O(n)$.

To prove that the number of strides obtained by this simple algorithm is linear,
we first define two notations, $a_{\max}(\omega)$ and $a_{\min}(\omega)$,
as follows.

\begin{definition}
\label{def:a_max}
 For a given orientation~$\omega \in [0, 2\pi]$,
 let~$a_{\max}(\omega) = \max_{0 \leq \theta \leq \omega} a(\theta)$
 and $a_{\min}(\omega) = \min_{\omega \leq \theta \leq 2\pi} a(\theta)$.
\end{definition}
In other words, $a_{max}(\omega)$ (\emph{resp.,}~$a_{\min}(\omega)$) is
the index of the furthest (\emph{resp.,} closest) edge of~$Q$
from the corner~$a_{0}$ along~$\partial Q$ in the counter-clockwise direction
among the edges visited by~$\atheta$ during the perfect rotation~$\PP[0, \omega]$
(\emph{resp.,} $\PP[\omega, 2\pi]$)
(consider two edges~$e_{1}$ and~$e_{n+1}$ are distinct).
Both~$a_{\max}(\omega)$ and~$a_{\min}(\omega)$ are
monotonically increasing; thus, by the definition of~$a(\theta)$,
we obtain~$a_{\max}(2\pi) - a_{\max}(0) = a_{\min}(2\pi) - a_{\min}(0) = n$.

We analogously define $b_{\max}(\omega)$, $b_{\min}(\omega)$,
$c_{\max}(\omega)$, $c_{\min}(\omega)$.
Based on Definition~\ref{def:a_max}, we introduce the following definition:
\begin{definition}
\label{def:E_max}
 For a given orientation~$\omega \in [0, 2\pi]$,
 let $\EE_{\max}(\omega)  = (a_{\max}(\omega),
 b_{\max}(\omega), c_{\max}(\omega))$
 and $\EE_{\min}(\omega) = (a_{\min}(\omega),
 b_{\min}(\omega), c_{\min}(\omega))$.
\end{definition}

From Corollary~\ref{cor:placement_ac} and Lemma~\ref{lem:placement_ar},
we observe that some specific pivots restrict the placement of $\Ptheta$.
Thus, based on Corollary~\ref{cor:placement_ac} and Lemma~\ref{lem:placement_ar},
we prove the following lemma about $\EE_{\min}(\omega)$:

\begin{corollary}
\label{cor:pivot_min}
 For a given step $\PP[\sigma, \sigma']$ with $\TT[\sigma, \sigma'] = a_{c}$ or $a_{r}$,
 $c_{\min}(\sigma) = c(\sigma)$ and $c_{\min}(\sigma') = c(\sigma')$.
 In particular, if $\sigma'$ is $2$-pivot when $\TT[\sigma, \sigma'] = a_{c}$, or
 if $c_{\sigma'}$ coincides with $q_{c(\sigma')}$ when $\TT[\sigma, \sigma'] = a_{r}$,
 then $c_{\min}(\omega)$ increases.
\end{corollary}
\begin{proof}
 Suppose that $\TT[\sigma, \sigma'] = a_{c}$ or $a_{r}$
 and $c_{\min}(\sigma') \neq c(\sigma')$.
 Then, there exists an orientation $\tau > \sigma'$ such that
 the corner $c_{\tau}$ moves backward
 and coincides with vertex $q_{c(\sigma')}$.
 Since $\tau > \sigma'$, $(1)$ the corner $a_{\tau}$ must lie
 on $\partial Q(a_{\sigma'}, c_{\sigma'})$ when $\TT[\sigma, \sigma'] = a_{c}$
 and $(2)$ the corner $b_{\tau}$ must lie
 on $\partial Q(b_{\sigma'}, c_{\sigma'})$ when $\TT[\sigma, \sigma'] = a_{r}$
 (otherwise, $c(\tau)  = c(\sigma) + n$).
 However, for $\theta > \tau$, if corner $\ctheta$ moves backward from $q_{c(\sigma')}$,
 $(1)$ corner $\atheta$ lies on $\partial Q(a_{\sigma'}, c_{\sigma'})$
 since it moves forward from $a_{\tau}$ and,
 $(2)$ corner $\btheta$ lies on $\partial Q(b_{\sigma}, c_{\sigma})$
  since it moves forward from $b_{\tau}$, respectively.
 This contradicts the definition of $\Ptheta$
 by Corollary~\ref{cor:placement_ac} and Lemma~\ref{lem:placement_ar}.
 Therefore, $c_{\min}(\sigma')  = c(\sigma')$ and
 and the same arguement applies to the pivot $\sigma$ by replacing $\sigma'$ with $\sigma$.
 
 If $\sigma'$ is $2$-pivot when $\TT[\sigma, \sigma'] = a_{c}$, or
 if $c_{\sigma'}$ coincides with $q_{c(\sigma')}$, this means that
 $c(\theta)$ increases at the pivot $\sigma'$.
 Then, $c_{\min}(\omega)$ increases at the pivot $\sigma'$
 since $c_{\min}(\omega) = c(\omega)$.
\end{proof}

\noindent Corollary~\ref{cor:pivot_min} remains valid
when $a, b, c$ are cyclcially rotated.

To track whether the current location of corner~$\atheta$
is the furthest from~$a_{0}$ along~$\partial Q$ in the counter-clockwise direction
during the perfect rotation~$\PP[0, \theta]$,
we use stack~$\mathcal{S}_{a}$ in the following manner.
Suppose that the type of step changes
from $f_{w}$, $a_{c}$, or~$c_{a}$ to~$a_{r}$ at a $3$-pivot~$\beta$.
Then, we mark the moment when corner~$\atheta$ begins to move backward
by pushing the $3$-pivot~$\beta$ onto stack~$\mathcal{S}_{a}$.
We pop this pivot from stack~$\mathcal{S}_{a}$
once corner~$\atheta$ lies on the counter-clockwise side of~$a_{\beta}$
along $\partial Q$ for some~$\theta > \beta$.

We also consider the two stacks~$\mathcal{S}_{b}$ and $\mathcal{S}_{c}$
in a cyclic manner. With three stacks~$\mathcal{S}_{a}$,
$\mathcal{S}_{b}$, and $\mathcal{S}_{c}$,
we prove the following lemma.
\begin{lemma}
\label{lem:prop_stack}
 Suppose that the type of step changes
 from~$f_{w}$, $a_{c}$, or~$c_{a}$ to~$a_{r}$ at a $3$-pivot~$\beta$;
 in other words, the $3$-pivot~$\beta$ is pushed onto stack~$\mathcal{S}_{a}$.
 Then, while the $3$-pivot~$\beta$ remains in stack~$\mathcal{S}_{a}$,
  for~$\theta > \beta$,
 \begin{itemize}
  \item[$(i)$] if corner~$\atheta$ lies on~$\partial Q$,
  corner~$\atheta$ must lie on~$\partial Q(c_{\beta}, a_{\beta})$; and
  if both corners~$\atheta$ and~$\ctheta$ lie on~$\partial Q$,
  corner~$\ctheta$ also lies on~$\partial Q(c_{\beta}, a_{\beta})$
  and the type of step after the $3$-pivot~$\beta$ cannot be $a_{c}$ or~$c_{a}$.
  \item[$(ii)$] if corner~$\atheta$ lies on~$\partial Q$,
  corner~$\btheta$ must lie on~$\partial Q(b_{\beta}, c_{\beta})$
  and the type of step after the $3$-pivot~$\beta$ cannot be $c_{r}$;
  in other words, corner~$\ctheta$ cannot move backward.
  \item[$(iii)$] stack $\mathcal{S}_{b}$ is empty at the $3$-pivot $\beta$.
  In particular, if a stride $\PP[\sigma, \sigma'']$ contains
  the $3$-pivot~$\beta \in [\sigma, \sigma'']$,
  then, $b_{\max}(\omega)$ increases after passing the stride.
 \end{itemize}
\end{lemma}
\begin{proof}
 $(i)$ For $\theta > \beta$, corner~$\ctheta$ cannot coincide with
 corner~$c_{\beta}$ because stack $\mathcal{S}_{a}$ is non-empty.
 If corner~$\ctheta$ lies on $\partial Q(b_{\beta}, c_{\beta})$ for $\theta > \beta$,
 then corner~$\btheta$ also lies on $\partial Q(b_{\beta}, c_{\beta})$
 (otherwise, $a(\theta) > a(\beta)$).
 However, the type of step changes to~$a_{r}$ at the $3$-pivot~$\beta$;
 thus, this contradicts the definition of $\Ptheta$ by Lemma~\ref{lem:placement_ar}.
 Consequently, if corner~$\ctheta$ lies on~$\partial Q$,
 then it must lie on~$\partial Q(c_{\beta}, a_{\theta})$,
 which implies that it lies on~$\partial Q(c_{\beta}, a_{\beta})$.
 Since corner~$\atheta$ lies on~$\partial Q(\ctheta, a_{\beta})$,
 it must lie on~$\partial Q(c_{\beta}, a_{\beta})$ as well,
 whenever corner~$\ctheta$ lies on~$\partial Q$.
 
 The remaining case is when corner~$\ctheta$ lies in the interior of~$Q$.
 Consider the moment that corner~$\ctheta$ departs from~$\partial Q$.
 By the previous result, both corners $\ctheta$ and $\atheta$
 must lie on~$\partial Q(c_{\beta}, a_{\beta})$.
 From this moment, corner~$\atheta$ continuously moves forward until
 corner~$\ctheta$ returns to $\partial Q$.
 Thus, corner $\atheta$ remains on~$\partial Q(c_{\beta}, a_{\beta})$. 
 
 If both corners $\atheta$ and~$\ctheta$ lie on $\partial Q$,
 they necessarily lie on $\partial Q(c_{\beta}, a_{\beta})$.
 Then, by Corollary~\ref{cor:placement_ac},
 the placement of $\Ptheta$ cannot lie on a step of type~$a_{c}$ or~$c_{a}$.
 
 \medskip
 $(ii)$ Since the type of step after the $3$-pivot $\beta$ cannot be $a_{c}$ or~$c_{a}$,
 corner~$\btheta$ must lie on~$\partial Q$.
 For $\theta > \beta$, corner~$\btheta$ must lie
 on $\partial Q(b_{\beta}, c_{\beta})$
 (otherwise $a(\theta) > a(\beta)$).
 By the convexity of~$Q$, we obtain
 $\varphi_{ab}(\theta) \leq \phi_{ab}(\beta) \leq \pi - \angle c$,
 which implies that corner~$\ctheta$ always moves forward.
 
 \medskip
 $(iii)$ When the $3$-pivot~$\beta$ is pushed onto stack~$\mathcal{S}_{a}$,
 either $(1)$ corner~$\btheta$ moves forward along~$\partial Q$
 until corner~$b_{\beta}$ coincides with vertex~$q_{b(\beta)}$, or
 $(2)$ corner~$\btheta$ moves through the interior of~$Q$
 and returns to~$\partial Q$ at edge~$e_{b(\beta)}$.
 In either case, we obtain $\angle a + \phi_{bc}(\beta) > \pi$;
 thus, by applying~$(ii)$ cyclically, stack~$\mathcal{S}_{b}$
 must be empty at the $3$-pivot $\beta$.
 
 In the former case $(1)$, the type of step changes
 from $f_{w}$ to $a_{r}$, which increases $b(\theta)$
 from $b(\beta) - 1$ to $b(\beta)$ at the $3$-pivot $\beta$;
 thus, $b_{\max}(\omega)$ also increases 
 since stack $\mathcal{S}_{b}$ is empty
 at the $3$-pivot $\beta$.
 
 In the latter case $(2)$, the type of step changes
 from $a_{c}$ or $c_{a}$ to $a_{r}$.
 Then, stack $\mathcal{S}_{b}$ can
 contain the $3$-pivot $\gamma$
 until corner $\btheta$ reaches corner $b_{\beta}$.
 We claim that $b(\beta) > b(\gamma)$,
 which implies that $b_{\max}(\omega)$ increases at the $3$-pivot $\beta$.
 
 Consider the largest $3$-pivot $\beta^{(0)} < \beta$
 where corner $\btheta$ departs from $\partial Q$.
 Since the type of step changes to $b_{r}$ at the $3$-pivot $\gamma$,
 we have $\beta^{(0)} > \gamma$.
 
 If $c_{\beta^{(0)}}$ lies on $\partial Q(b_{\gamma}, c_{\gamma})$,
 by $(ii)$, $b(\beta^{(0)}) \geq b(\gamma)$;
 in other words, stack $\mathcal{S}_{b}$ is already empty
 at the $3$-pivot $\beta^{(0)}$.
 Since the type of step  changes from $a_{c}$ or $c_{a}$ to $a_{r}$,
 we have $b(\beta) > b(\beta^{(0)})$ by Lemma~\ref{lem:type_abba},
 meaning that $b(\beta) > b(\gamma)$.
 
 Otherwise, $c(\beta^{(0)}) \geq c(\gamma)$.
 During the perfect rotation $\PP[\beta^{(0)}, \beta]$,
 corner $\ctheta$ always moves forward, and thus, $c(\beta) \geq c(\beta^{(0)})$.
 If $b(\beta) = b(\gamma)$, then
 we have $\phi_{bc}(\beta) \leq \phi_{bc}(\gamma)$.
 However, the type of step changes to $b_{r}$ at the pivot $\gamma$ and
 changes to $a_{r}$ at the pivot $\beta$,
 we obtain $\phi_{bc}(\gamma) \leq \pi - \angle a < \phi_{bc}(\beta)$,
 which contradicts with the previous inequality.
 Therefore, $b(\beta) > b(\gamma)$ which implies that
 $b_{\max}(\omega)$ increases. 
 
 If a stride $\PP[\sigma, \sigma'']$ contains
 the $3$-pivot~$\beta \in (\sigma, \sigma'']$,
 then it cannot contain another $3$-pivot~$\gamma \in (\beta, \sigma'')$
 where the type of step changes from~$f_{w}$, $b_{a}$, or~$a_{b}$ to~$b_{r}$
 since both $b(\theta)$ and $c(\theta)$ of~$\Etheta$
 cannot be updated during a single stride.
 Therefore, if any $3$-pivot is pushed onto stack $\mathcal{S}_{a}$ during a stride,
 $b_{\max}(\omega)$ increases after the stride.
\end{proof}
\noindent Lemma~\ref{lem:prop_stack} remains valid
when $a, b, c$ are cyclically rotated.
In general, the type of step can be restricted
depending on whether stacks~$\mathcal{S}_{a}$, $\mathcal{S}_{b}$,
and~$\mathcal{S}_{c}$ are empty.
If any $3$-pivot is pushed onto one of stacks~$\mathcal{S}_{a}$,
$\mathcal{S}_{b}$, or~$\mathcal{S}_{c}$ during a stride,
then at least one element of $\EE_{\max}(\omega)$
increases after the stride.

\begin{theorem}
\label{thm:linear}
  The global perfect rotation~$\PP[0, 2\pi]$ can be covered by a
  linear number of strides.
\end{theorem}
\begin{proof}
 Without loss of generality, suppose that orientation~$0$ is a pivot of~$\Lambda$
 and all three stacks~$\mathcal{S}_{a}$, $\mathcal{S}_{b}$, $\mathcal{S}_{c}$
 are initially empty. We then sequentially apply
 Lemma~\ref{lem:take_stride} from $\PP(0)$
 until the global perfect rotation is entirely covered by strides.
 We claim that the total number of these strides is at most linear.
 
 While all three stacks~$\mathcal{S}_{a}$, $\mathcal{S}_{b}$, $\mathcal{S}_{c}$ are empty,
 if a stride contains no steps of type~$a_{r}$, $b_{r}$, or~$c_{r}$,
 then at least one element of $\EE_{\max}(\omega)$ increases after the stride.
 Thus, the number of such strides is at most linear.
 
 Without loss of generality,  we assume that $\PP[\xi_{1}, \xi_{1}'']$ is
 the first stride after pivot $0$ that contains a $3$-pivot $\beta^{(1)} \in (\xi_{1}, \xi_{1}'']$
 where the type of step changes from $f_{w}$, $a_{c}$, or~$c_{a}$ to $a_{r}$.
 Then the $3$-pivot $\beta^{(1)}$ is pushed onto stack $\mathcal{S}_{a}$
 during this stride and, by part~$(iii)$ of Lemma~\ref{lem:prop_stack},
 $b_{\max}(\omega)$ increases after passing the stride~$\PP[\xi_{1}, \xi_{1}'']$
 since stack $\mathcal{S}_{b}$ is empty.
 
 While the $3$-pivot $\beta^{(1)}$ remains in stack $\mathcal{S}_{a}$,
 by Lemma~\ref{lem:prop_stack},
 a stride cannot contain steps of type~$a_{c}$, $c_{a}$, or~$c_{r}$ and
 no pivot can be pushed onto stack~$\mathcal{S}_{c}$.
 Consequently, either the $3$-pivot $\beta^{(1)}$ will be popped from stack $\mathcal{S}_{a}$
 or another $3$-pivot $\gamma^{(1)}$ will be pushed onto stack $\mathcal{S}_{b}$.
 Before either of these two events occurs,
 we examine how the strides affect $\EE_{\max}(\omega)$ or $\EE_{\min}(\omega)$;
 that is, we consider the strides while only stack $\mathcal{S}_{a}$ is non-empty.
 
 From Lemma~\ref{lem:prop_stride}, it is enough to consider the strides
 that contain no steps of type~$a_{c}$, $c_{a}$, $c_{r}$, or~$b_{r}$.
 For a stride $\PP[\sigma, \sigma'']$
 while only stack $\mathcal{S}_{a}$ is non-empty,
 at least one of~$b_{\max}(\omega)$, $c_{\max}(\omega)$,
 or~$a_{\min}(\omega)$ increases
 after passing the stride as follows:
 
 \begin{enumerate}
  \item From $\TT[f_{w}] = \{f_{w}\} \cup 
  \{f_{w}s \mid s \in \TT_{a} \cup \TT_{b} \cup \TT_{c}\}$,
  where the set $\TT_{a} = \{\varepsilon, a_{c}, c_{r}, c_{r}a_{c}, c_{r}c_{a}\}$,
  the set $\TT_{b} = \{\varepsilon, b_{a}, a_{r}, a_{r}b_{a}, a_{r}a_{b}\}$, and
  the set $\TT_{c} = \{\varepsilon, c_{b}, b_{r}, b_{r}c_{b}, b_{r}b_{c}\}$:
  \begin{enumerate}
   \item when $a(\theta)$ is updated first, $\TT[\sigma, \sigma''] = f_{w}$:
   either $b(\sigma'') = b(\sigma) + 1$ or $c(\sigma'') = c(\sigma) + 1$;
   thus, either $b_{\max}(\omega)$ or $c_{\max}(\omega)$ increases.
   \item when either $b(\theta)$ or $c(\theta)$ is updated first,
   then $b_{\max}(\omega)$ or $c_{\max}(\omega)$ increases
   since $b(\sigma'') > b(\sigma)$ or $c(\sigma'') > c(\sigma)$, respectively.
  \end{enumerate}
  
  \item From $\TT[b_{a}] = \{b_{a}, b_{a}b_{c}, b_{a}b_{r}\}$:
  \begin{enumerate}
   \item $\TT[\sigma, \sigma''] = b_{a}$:
   since $a_{\sigma''}$ coincides with a vertex of $Q$,
   by Corollary~\ref{cor:pivot_min}, $a_{\min}(\omega)$ increases.
   \item $\TT[\sigma, \sigma''] = b_{a}b_{c}$:
   $c_{\max}(\omega)$ increases since $c(\sigma'') > c(\sigma)$.
  \end{enumerate}
  
  \item From $\TT[c_{b}] = \{c_{b}, c_{b}c_{a}, c_{b}c_{r}\}$:
  $\TT[\sigma, \sigma''] = c_{b}$ and thus, $b_{\max}(\omega)$ increases
  since $b(\sigma'') > b(\sigma)$.
  
  \item From $\TT[a_{b}] = \{a_{b}s \mid s \in \TT[a_{c}] \cup \TT[f_{w}]
  \cup (\TT_{c} \setminus \{c_{b}\})\}$:
  \begin{enumerate}
   \item $\TT[\sigma, \sigma''] = a_{b}$: $b_{\max}(\omega)$ increases
   since $b(\sigma'') > b(\sigma)$.
   \item $\TT[\sigma, \sigma''] \in \{a_{b}s \mid s \in \TT[f_{w}]\}$:
   by the analysis of $\TT[f_{w}]$,
   either $b(\sigma'') > b(\sigma)$ or $c(\sigma'') > c(\sigma)$.
  \end{enumerate}
  
  \item From $\TT[b_{c}] = \{b_{c}s \mid s \in \TT[b_{a}] \cup \TT[f_{w}]
  \cup (\TT_{a} \setminus \{a_{c}\})\}$:
  \begin{enumerate}
   \item $\TT[\sigma, \sigma''] = b_{c}$: $c_{\max}(\omega)$ increases
   since $c(\sigma'') > c(\sigma)$.
   \item $\TT[\sigma, \sigma''] \in \{b_{c}s \mid s \in \TT[b_{a}]\}$:
   by the analysis of $\TT[b_{a}]$, either $a_{\min}(\omega)$ increases or
   $c(\sigma'') > c(\sigma)$.
   \item $\TT[\sigma, \sigma''] \in \{b_{c}s \mid s \in \TT[f_{w}]\}$:
   by the analysis of $\TT[f_{w}]$,
   either $b(\sigma'') > b(\sigma)$ or $c(\sigma'') > c(\sigma)$. 
  \end{enumerate}
  
  \item From $\TT[a_{r}] =  \{a_{r}b_{a}, a_{r}a_{b}a_{c}\}
  \cup \{a_{r}s \mid s \in \TT_{c} \setminus \{c_{b}\}\} 
  \cup \{a_{r}a_{b}s \mid s \in \TT_{c} \setminus \{c_{b}\}\}
  \cup \{a_{r}f_{w}s \mid s \in~\TT_{c}\}
  \cup \{a_{r}a_{b}f_{w}s \mid s \in \TT_{c}\}$:
  \begin{enumerate}
   \item $\TT[\sigma, \sigma''] = a_{r}$ or $a_{r}a_{b}$:
   $b_{\max}(\omega)$ or $c_{\max}(\omega)$ increases
   since $b(\sigma'') > b(\sigma)$ or $c(\sigma'') > c(\sigma)$.
   \item $\TT[\sigma, \sigma''] = a_{r}b_{a}$:
   $b_{\max}(\omega)$ increases since $b(\sigma'') > b(\sigma)$.
   \item $\TT[\sigma, \sigma''] \in
   \{a_{r}f_{w}, a_{r}f_{w}c_{b}, a_{r}a_{b}f_{w}, a_{r}a_{b}f_{w}c_{b}\}$:
   $c_{\max}(\omega)$ increases since $c(\sigma'') > c(\sigma)$.
  \end{enumerate}
 \end{enumerate}
 
 \noindent This implies that the number of strides
 while only stack $\mathcal{S}_{a}$ is non-empty
 is at most linear.
 Since this observation remains valid when $a, b, c$ are cyclically rotated,
 in general, the number of strides while only one of the stacks
 $\mathcal{S}_{a}$, $\mathcal{S}_{b}$, $\mathcal{S}_{c}$ is non-empty
 is also at most linear.
 
 \medskip
 Suppose that the $3$-pivot $\beta^{(1)}$ is popped first from stack $\mathcal{S}_{a}$
 at the orientation $\beta^{\ast} \in (\xi_{2}, \xi_{2}'']$
 where $\PP[\xi_{2}, \xi_{2}'']$ is a stride.
 If this stride contains type $b_{r}$ or $c_{r}$ step,
 some $3$-pivot is pushed onto stack $\mathcal{S}_{b}$ or $\mathcal{S}_{c}$
 during the stride since there exists no $b_{r}$ or $c_{r}$ steps
 within the perfect rotation $\PP[\xi_{2}, \beta^{\ast}]$.
 Thus, $b_{\max}(\omega)$ or $c_{\max}(\omega)$ increases
 after passing the stride by part $(iii)$ of Lemma~\ref{lem:prop_stack}.
 
 If the stride $\PP[\xi_{2}, \xi_{2}'']$ is
 a step of type $a_{b}$, $b_{a}$, $b_{c}$, or $c_{b}$,
 one element of $\EE_{max}(\omega)$ increases after passing the stride.
 Otherwise, $\EE(\xi_{2}'')$ differs from $\EE(\xi_{2})$
 in at least two elements; thus,
 $b_{\max}(\omega)$ or $c_{\max}(\omega)$ increases.
 At the orientation $\beta^{\ast}$,
 all three stacks $\mathcal{S}_{a}$, $\mathcal{S}_{b}$, $\mathcal{S}_{c}$
 become empty, which restores the initial configuration.
 Thus, we can consider the pivot $\xi_{2}$ as the orientation $0$.
 
 We now consider the case when another $3$-pivot $\gamma^{(1)}$
 is pushed onto stack $\mathcal{S}_{b}$ after the $3$-pivot $\beta^{(1)}$.
  Suppose that $\PP[\eta_{1}, \eta_{1}'']$ is the first stride after pivot $\xi_{1}''$
 that contains a $3$-pivot $\gamma^{(1)} \in (\eta_{1}, \eta_{1}'']$
 where the type of step changes from $f_{w}$, $a_{b}$, or~$b_{a}$ to $b_{r}$.
 Then the $3$-pivot $\gamma^{(1)}$ is pushed onto stack $\mathcal{S}_{b}$
 during this stride and, by part $(iii)$ of Lemma~\ref{lem:prop_stack},
 $c_{\max}(\omega)$ increases after passing the stride $\PP[\eta_{1}, \eta_{1}'']$
 since stack $\mathcal{S}_{c}$ is empty.
 
 While two stacks~$\mathcal{S}_{a}$ and $\mathcal{S}_{b}$ are non-empty,
 a stride can contain only steps of type~$f_{w}$, $b_{c}$, $c_{b}$, or~$b_{r}$
 and no pivot can be pushed onto stacks $\mathcal{S}_{c}$ or $\mathcal{S}_{a}$.
 Then we first enumerate all possible type of strides
 while two stacks~$\mathcal{S}_{a}$ and $\mathcal{S}_{b}$ are non-empty as follows:
  \begin{enumerate}
  \item From $\TT[f_{w}] = \{f_{w}\} \cup 
  \{f_{w}s \mid s \in \TT_{a} \cup \TT_{b} \cup \TT_{c}\}$:
  the possible types $\TT[\sigma, \sigma''] \in \{f_{w}s \mid s \in \TT_{c}\}$.
  \item From $\TT[c_{b}] = \{c_{b}, c_{b}c_{a}, c_{b}c_{r}\}$:
  the possible type $\TT[\sigma, \sigma''] = c_{b}$.
  \item From $\TT[b_{c}] = \{b_{c}s \mid s \in \TT[b_{a}] \cup \TT[f_{w}]
  \cup (\TT_{a} \setminus \{a_{c}\})\}$:
  the possible types $\TT[\sigma, \sigma''] \in \{b_{c}\} \cup
  \{b_{c}f_{w}s \mid s \in \TT_{c}\}$
  \item From $\TT[b_{r}] =  \{b_{r}c_{b}, b_{r}b_{c}b_{a}\}
  \cup \{b_{r}s \mid s \in \TT_{a} \setminus \{a_{c}\}\} 
  \cup \{b_{r}b_{c}s \mid s \in \TT_{a} \setminus \{a_{c}\}\}
  \cup \{b_{r}f_{w}s \mid s \in \TT_{a}\}
  \cup \{b_{r}b_{c}f_{w}s \mid s \in \TT_{a}\}$:
  the possible types $\TT[\sigma, \sigma''] \in \{b_{r}, b_{r}b_{c},
  b_{r}c_{b}, b_{r}f_{w}, b_{r}b_{c}f_{w}\}$.
 \end{enumerate}

 By Lemma~\ref{lem:prop_stride}, $c_{\max}(\omega)$ increases
 after passing a stride with $\TT[\sigma, \sigma''] \in 
 \{f_{w}s \mid s \in \TT_{c} \setminus \{\varepsilon\}\} \cup
 \{b_{c}, b_{r}c_{b}\} \cup
 \{b_{c}f_{w}s \mid s \in \TT_{c}\setminus \{\varepsilon\}\} $,
 since $c(\sigma'') > c(\sigma)$.
 Similarly, after passing a stride with $\TT[\sigma, \sigma''] = c_{b}$,
 by Corollary~\ref{cor:pivot_min}, $b_{\min}(\omega)$ increases.
 By Lemma~\ref{lem:prop_stride} again, after passing
 a stride with one of the remaining types $\TT[\sigma, \sigma''] \in
 \{f_{w}, b_{r}, b_{r}b_{c}, b_{r}f_{w}, b_{c}f_{w}, b_{r}b_{c}f_{w}\}$,
 either $a(\sigma'') > a(\sigma)$ or $c(\sigma'') > c(\sigma)$  is satisfied.
 Since $c_{\max}(\omega)$ increases if $c(\sigma'') > c(\sigma)$,
 we assume that each stride with $\TT[\sigma, \sigma''] \in
 \{f_{w}, b_{r}, b_{r}b_{c}, b_{r}f_{w}, b_{c}f_{w}, b_{r}b_{c}f_{w}\}$
 satisfies $a(\sigma'') > a(\sigma)$.
 
 Let $\PP[\zeta_{1}, \zeta_{1}'']$ be the last stride such that
 $(i)$ $\zeta_{1} \geq \eta_{1}$ and the $3$-pivot~$\beta^{(1)}$
 remains in stack~$\mathcal{S}_{a}$ at the pivot~$\zeta_{1}$,
 $(ii)$ the stride~$\PP[\zeta_{1}, \zeta_{1}'']$ contains type~$b_{r}$ step,
 and $(iii)$ the perfect rotation~$\PP[\eta_{1}, \zeta_{1}]$
 contains no type~$a_{r}$ step
 (where $\eta_{1} = \zeta_{1}$ and $\eta_{1}'' = \zeta_{1}''$ may hold).
 During the perfect rotation $\PP[\eta_{1}, \zeta_{1}]$,
 stack $\mathcal{S}_{b}$ can alternately become empty and non-empty.
 
 In particular, let $\alpha^{(1)} \in (\zeta_{1}, \zeta_{1}'']$ be the $3$-pivot
 where the type of step changes from $b_{r}$ to another type.
 By Corollary~\ref{cor:pivot_min}, $a_{\min}(\alpha^{(1)}) = a(\alpha^{(1)})$
 and by Lemma~\ref{lem:prop_stride}, the stride $\PP[\zeta_{1}, \zeta_{1}'']$
 cannot contain type $a_{r}$ step since it contains type $b_{r}$ step;
 in other words, corner $\atheta$ moves forward
 during the perfect rotation $\PP[\eta_{1}, \alpha^{(1)}]$.
 Therefore, each stride with $\TT[\sigma, \sigma''] \in
 \{f_{w}, b_{r}, b_{r}b_{c}, b_{r}f_{w}, b_{c}f_{w}, b_{r}b_{c}f_{w}\}$
 that appears within the perfect rotation $\PP[\eta_{1}, \alpha^{(1)}]$ 
 while two stacks $\mathcal{S}_{a}$ and $\mathcal{S}_{b}$ are non-empty
 increases $a_{\min}(\omega)$ by at least one.
 
 Additionally, if stack $\mathcal{S}_{b}$ becomes empty
 while $\mathcal{S}_{a}$ remains non-empty
 after passing a stride that appears within the perfect rotation $\PP[\eta_{1}, \zeta_{1}]$,
 then at least one of $b_{\max}(\omega)$, $c_{\max}(\omega)$,
 or $a_{\min}(\omega)$ increases after passing the stride.
 We have already shown that at least one element of
 $\EE_{\max}(\omega)$ or $\EE_{\min}(\omega)$ increases after passing
 either a stride that occurs while only stack $\mathcal{S}_{a}$ is non-empty
 or a stride that pushes a $3$-pivot onto stack $\mathcal{S}_{b}$.
 Consequently, every stride that appears within
 the perfect rotation $\PP[\eta_{1}, \zeta_{1}]$
 increases at least one element of
 $\EE_{\max}(\omega)$ or $\EE_{\min}(\omega)$.
 
 From the pivot $\zeta_{1}$, we now take strides until one of two stacks
 $\mathcal{S}_{a}$ or $\mathcal{S}_{b}$ becomes empty.
 From now on, while two stacks $\mathcal{S}_{a}$ and $\mathcal{S}_{b}$
 are non-emtpy, we will focus on the strides which
 cannot affect the elements of $\EE_{\max}(\omega)$ or $\EE_{\min}(\omega)$.
 Then, by the definition of stride $\PP[\zeta_{1}, \zeta_{1}'']$,
 it is enough to consider the strides with $\TT[\sigma, \sigma''] \in 
 \{f_{w}, b_{c}f_{w}\}$ after the stride $\PP[\zeta_{1}, \zeta_{1}'']$.
 Since the type of the last step of these strides is $f_{w}$,
 we assume that both $a(\sigma'') > a(\sigma)$ and
 $b(\sigma'') > b(\sigma)$ hold
 ($c_{\max}(\omega)$ increases if $c(\sigma'') > c(\sigma)$).
 
 Suppose that $\gamma^{\ast}$ be the bottom pivot
 in stack $\mathcal{S}_{b}$ after the $3$-pivot $\alpha^{(1)}$.
 Then, let $\PP[\chi_{1}, \chi_{1}'']$ be the stride
 with $\TT[\chi_{1}, \chi_{1}''] = f_{w}$ or $b_{c}f_{w}$ such that
 $\chi_{1} \geq \zeta^{(1)}$ and
 $(1)$ stack $\mathcal{S}_{a}$ becomes empty first after passing the stride,
 or $(2)$ the last stride where $b_{\chi_{1}''}$ lies on
 clockwise side of $b_{\gamma^{\ast}}$.
 By the definitions, after passing each stride within $\PP[\alpha^{(1)}, \chi_{1}'']$,
 $a(\theta)$ increases (the stride $\PP[\zeta_{1}, \zeta_{1}'']$ increases
 $a(\theta)$ or $a_{\min}(\omega)$); thus, there are at most
 $a(\chi_{1}'') - a(\alpha^{(1)})$ number of strides
 which cannot affect the elements of $\EE_{\max}(\omega)$ or $\EE_{\min}(\omega)$.
 
 Consider the smallest pivot $\xi_{3} > \chi_{1}''$ such that the
 such that the $3$-pivot $\beta^{(2)} \in (\xi_{3}, \xi_{3}'']$ is pushed
 onto stack $\mathcal{S}_{a}$ during the stride $\PP[\xi_{3}, \xi_{3}'']$.
 Then, corner $\atheta$ moves forward during
 the perfect rotation $\PP[\chi_{1}'', \beta^{(2)}]$,
 i.e. $a(\chi_{1}'') \leq a(\beta^{(2)})$.
 Since two $3$-pivots $\gamma^{\ast}$ and $\beta^{(2)}$
 satisfy $\gamma^{\ast} < \beta^{(2)}$ and
 $\phi_{bc}(\gamma^{\ast}) \leq \pi - \angle a
 < \phi_{bc}(\beta^{(2)})$, by Lemma~\ref{lem:placement_ar},
 we have $b(\gamma^{\ast}) < b(\beta^{(2)})$.
 
 Define the strides $\PP[\eta_{2}, \eta_{2}'']$, $\PP[\zeta_{2}, \zeta_{2}'']$,
 and $\PP[\chi_{2}, \chi_{2}'']$ and
 the pivots $\gamma^{(2)}$ and $\alpha^{(2)}$, analogously.
 Then, every stride that appear within the perfect rotaion $\PP[\eta_{2}, \zeta_{2}]$
 increases at least one element of $\EE_{\max}(\omega)$ or $\EE_{\min}(\omega)$.
 Further, there are at most $a(\chi_{2}'') - a(\alpha^{(2)})$ number of strides
 which cannot affect the elements of $\EE_{\max}(\omega)$ or $\EE_{\min}(\omega)$.
 We finally claim that $a_{\alpha^{(2)}}$ lies on the counter-clockwise side of
 $a_{\chi_{1}''}$.
 
 If $b_{\alpha^{(2)}}$ lies in the clockwise side of $b_{\beta^{(2)}}$,
 then $a_{\alpha^{(2)}}$ lies in the counter-clockwise side of $a_{\beta^{(2)}}$
 since $\alpha^{(2)} < \beta^{(2)}$.
 Thus, $a_{\alpha^{(2)}}$ lies in the counter-clockwise side of $a_{\chi_{1}''}$
 
 Otherwise, $b(\alpha^{(2)}) \geq b(\beta^{(2)}) > b(\gamma^{\ast}) > b(\chi_{1}'')$,
 in other words, the corners $b_{\chi_{1}''}$, $b_{\gamma^{\ast}}$, $b_{\alpha^{(2)}}$
 appears in counter-clockwise order along $\partial Q$.
 Since stack $\mathcal{S}_{a}$ is non-empty
 during the perfect rotation $\PP[\gamma^{\ast}, \chi_{1}'']$
 and stack $\mathcal{S}_{c}$ is empty at the pivot $\gamma^{(2)}$,
 we have $c(\gamma^{\ast}) \leq c(\chi_{1}'') < c(\alpha^{(2)})$,
 in other words, $c_{\gamma^{\ast}}$, $c_{\chi_{1}''}$, $c_{\alpha^{(2)}}$ appears
 in counter-clockwise order along $\partial Q$.
 Then, by Corollary~\ref{cor:two_triangles},
 $\phi(\overrightarrow{c_{\chi_{1}''}c_{\alpha^{(2)}}},
 \overrightarrow{b_{\alpha^{(2)}}b_{\chi_{1}''}}) < 
 \phi(\overrightarrow{c_{\gamma^{\ast}}c_{\alpha^{(2)}}},
 \overrightarrow{b_{\alpha^{(2)}}b_{\gamma^{\ast}}}) \leq \pi - \angle a$,
 which implies that $a_{\alpha^{(2)}}$ lies on the counter-clockwise side of
 $a_{\chi_{1}''}$.
 Therefore, the interval from $a_{\alpha^{(1)}}$ to $a_{\chi_{1}''}$ along $\partial Q$
 is disjoint with the interval from $a_{\alpha^{(2)}}$ to $a_{\chi_{2}''}$ along $\partial Q$.
 This means that the number of strides which cannot affect the elements of 
 $\EE_{\max}(\omega)$ or $\EE_{\min}(\omega)$
 while two stacks $\mathcal{S}_{a}$ and $\mathcal{S}_{b}$ are non-empty is linear.
 We apply this result by permuting $a, b, c$, cyclially.
 
 Consequently, the number of strides that entirely cover the global perfect rotation is linear. 
\end{proof}

\section{Conclusion}

By Theorem~\ref{thm:linear}, we can partition the global perfect rotation
into a linear number of strides.
As we have seen in Lemma~\ref{lem:take_stride},
each stride consists of at most three chains and two steps
of one of types $a_{b}$, $b_{a}$, $b_{c}$, $c_{b}$, $c_{a}$, or $a_{c}$.
Therefore, this proves Theorem~\ref{thm:main}.

Since a linear number of strides are obtained in counter-clockwise order
with respect to angle,
a linear number of chains and steps are also sorted in counter-clockwise order
with repsect to angle.
Thus, based on the two pivots that define each chain or each step,
for a given angle $\psi$, we can find a chain or a step which contains $\PP(\psi)$
in $O(\log n)$ time using binary search.

If a step contains $\PP(\psi)$, we can easily compute the placement of $\PP(\psi)$.
Otherwise, if a chain contains $\PP(\psi)$,
only one element of $\Etheta$ can be changed continuously during the chain.
Thus, we can present four corner-edge incidences of all pivots in the chain
with $O(1)$ space.
Since we know four corner-edge incidences of each pivot,
we also can compute the pivot $\sigma$ and the placement of $\PP(\sigma)$.

Then, there are at most $n$ pivots in each chain,
so we can find a step of the chain which contains $\PP(\psi)$
in $O(\log n)$ time using binary search
(computing the placement of mid-pivot, iteratively).
Therefore, for a given angle $\psi \in [0, 2\pi]$,
we can query the placement of $\PP(\psi)$ in $O(\log n)$ time
with a linear numer of chains and steps.

Finally, for each chain $\PP[\sigma^{(1)}, \sigma^{(2)}]$,
we can compute the number of pivots in chain by comparing
$\EE(\sigma^{(2)})$ with $\EE(\sigma^{(1)})$ in constant time.
Thus, we can compute the complexity of silhouette $\Lambda$
in $O(n \log n)$ time.
Since computing a step takes $O(\log n)$ time
by Corollary~\ref{cor:moving_dists} and Lemma~\ref{lem:take_stride},
we can solve the problems $(2)$ and $(3)$ which finds
the maximum and minimum of $\lambda(\theta)$ for $\theta \in [0, 2\pi]$
in $O(k \log n)$ time by taking steps from $\PP(0)$ to $\PP(2\pi)$,
where $k$ is the complexity of silhouette $\Lambda$
(in most cases, the complexity $k$ is linear unless $\omega(1)$ number of vertices of $Q$
are highly concentrated within a very narrow interval of $\partial Q$).

\newpage

\bibliography{paper}

@inproceedings{chazelle1983polygon,
  title={The polygon containment problem},
  author={Chazelle, Bernard},
  booktitle={Advances in Computing Research},
  editor={F. P. Preparata},
  publisher={JAI Press},
  pages={1--33},
  year={1983}
}

@article{bae2018reverse,
  title = {The reverse {Kakeya} problem},
  author = {Sang Won Bae and Sergio Cabello and Otfried Cheong and Yoonsung Choi and Fabian Stehn and Sang Duk Yoon},
  pages = {75--84},
  volume = {21},
  journal = {Advances in Geometry},
  doi = {10.1515/advgeom-2020-0030},
  year = {2021},
}

@inproceedings{chan2024convex,
  title={Convex Polygon Containment: Improving Quadratic to Near Linear Time},
  author={Chan, Timothy M and Hair, Isaac M},
  booktitle={40th International Symposium on Computational Geometry (SoCG 2024)},
  pages={34--1},
  year={2024},
  organization={Schloss Dagstuhl--Leibniz-Zentrum f{\"u}r Informatik}
}

@article{sharir1994external,
  title={External polygon containment problems},
  author={Sharir, Micha and Toledo, Sivan},
  journal={Computational Geometry},
  volume={4},
  pages={99--118},
  year={1994},
  publisher={Elsevier}
}

@article{agarwal1998largest,
  title={Largest placement of one convex polygon inside another},
  author={Agarwal, Pankaj K and Amenta, Nina and Sharir, Micha},
  journal={Discrete \& Computational Geometry},
  volume={19},
  pages={95--104},
  year={1998},
  publisher={Springer}
}

@article{lee2021largest,
  title={Largest triangles in a polygon},
  author={Lee, Seungjun and Eom, Taekang and Ahn, Hee-Kap},
  journal={Computational Geometry},
  volume={98},
  pages={101792},
  year={2021},
  publisher={Elsevier}
}

@inproceedings{depano1987finding,
  title={Finding largest inscribed equilateral triangles and squares},
  author={DePano, A and Ke, Yan and O’Rourke, Joseph},
  booktitle={Proc. 25th Allerton Conf. Commun. Control Comput},
  pages={869--878},
  year={1987},
  organization={Univ. Illinois}
}

@book{baird,
author = {Baird, Henry S.},
title = {Model-based image matching using location},
year = {1985},
isbn = {0262022206},
publisher = {Massachusetts Institute of Technology},
address = {USA}
}

@article{pal1921,
 author               = {G. P\'al},
 journal              = {Math. Ann.},
 pages                = {311--319},
 title                = {Ein {M}inimumproblem f{\"{u}}r {O}vale},
 volume               = {83},
 year                 = {1921},
 }
\end{document}